%% file: 2163_4sissa.tex
\documentclass[twocolumn]{aa}
\usepackage{graphicx}
\usepackage{txfonts}
\newcommand{\srm}{\scriptscriptstyle \rm}
\newcommand{\bgal}{b_{\srm II}}

\begin{document}

   \title{New optical polarization measurements of quasi-stellar
   objects. The data \thanks{Based on observations collected at the
   European Southern Observatory (ESO, La Silla and
   Paranal)}$^{\rm{,}}$ \thanks{Table~4 is only available
   in electronic form at CDS via anonymous ftp to {\texttt
   {cdsarc.u-strasbg.fr (130.79.128.5)}} or {\texttt
   {http://cdsweb.u-strasbg.fr/cgi-bin/qcat?J/A+A/433/757}}}}

   \author{D. Sluse\inst{1,2}, D. Hutsem\'ekers
    \inst{1}$^{\rm{,}}$\thanks{Chercheur qualifi\'e du F.N.R.S. (Belgique)},
    H. Lamy \inst{3}, R. Cabanac \inst{4,5}, H. Quintana\inst{5}}

    \offprints{sluse@astro.ulg.ac.be}

    \institute{Institut d'Astrophysique et de G\'eophysique,
    Universit\'e de Li\`ege, All\'ee du 6 Ao\^ut 17, B5C, B-4000 Li\`ege, 
    Belgium \and European Southern Observatory,
    Alonso de Cordova 3107, Santiago 19, Chile \and BIRA-IASB, Avenue
    Circulaire 3, B-1180 Bruxelles, Belgium \and Canada France Hawaii
    Telescope, 65-1238 Mamalahoa Highway, Kamuela, Hawaii 96743, USA
    \and Departamento de Astronom{\'\i}a y Astrof{\'\i}sica, Pontificia 
     Universidad Cat\'olica de Chile, Casilla 306, Santiago 22, Chile}

   \date{Received: 12 October 2004; Accepted: 15 December 2004}

   \abstract{New linear polarization measurements (mainly in the V
   band) are presented for 203 quasi-stellar objects (QSOs). The
   sample is made up of 94 QSOs located in the North Galactic Pole
   (NGP) region and of 109 QSOs in the South Galactic Pole (SGP)
   region. First time measurements have been obtained for 184
   QSOs. Among them, 109 known radio-emitters, 42 known Broad
   Absorption Line (BAL) QSOs, and 1 gravitationally lensed
   quasi-stellar object. We found high polarization levels (p $>$ 3\%)
   for 12 QSOs, including the BAL QSO \object{SDSS~J1409+0048}. For 10
   objects, measurements obtained at different epochs do exist. Two of
   them show evidence for variability: the highly polarized BL Lac
   candidate \object{PKS~1216-010} and the radio source
   \object{PKS~1222+037}.

   }

    \titlerunning{New optical polarization measurements of quasi-stellar
    objects} \authorrunning{Sluse D. et al.}

   \maketitle
%

\section{Introduction}
\label{sec:intro}

Based on a large sample of quasi-stellar objects with measured optical
polarization, Hutsem\'ekers ({\cite{HUT98a}}) discovered that there
exists regions in the sky where the QSO polarization vectors appear
concentrated along preferential directions, on scales up to $\sim$ 1
Gpc. New data enabled Hutsem\'ekers \& Lamy (\cite{HUT01}) to confirm
this effect. In order to obtain an accurate and complete description
of this intriguing phenomenon, new QSO polarization measurements are
badly needed. The present paper provides a new set of polarimetric
data for 94 QSOs located in the North Galactic Cap and for 109 QSOs
located in the South Galactic Pole region, with details on the
observations, data reduction, and measurements. The comprehensive
analysis and the interpretation of the full sample will be reported
elsewhere.

This sample may be used for a variety of other studies such as
investigating the relation between the optical polarization properties
of QSOs and their optical spectra (e.g. Stockman et
al. {\cite{STO84}}) or their radio properties (e.g. Berriman et
al. {\cite{BER90}}, Rusk {\cite{RUS90}}, Visvanathan \& Wills
{\cite{VIS98}}). It also enables one to study the polarization
properties of sub-classes of QSOs such as Broad Absorption Line (BAL)
quasi-stellar objects (e.g. Hutsem\'ekers et al. {\cite{HUT98b}},
Schmidt \& Hines {\cite{SCH99}}, Lamy \& Hutsem\'ekers {\cite{LAM04}})
or near-infrared selected QSOs (e.g. Smith et al. {\cite{SMI02}}).

\section {The observations}

The polarimetric observations were carried out during 5 runs at the
European Southern Observatory, La Silla, in August 2000, March 2002,
May 2002, August 2003 and October 2003, using the 3.6m telescope
equipped with EFOSC2. Two additional objects were observed on April
21, 2002 with EFOSC2 and three objects on February 25, 2003 in service
mode with the VLT UT1 equipped with the FORS1 camera. The CCD\#40
mounted on EFOSC2 is a 2k$\times$2k CCD with a pixel size of 15 $\mu$m
corresponding to 0.158\arcsec\ on the sky in the 1$\times$1 binning
mode. The standard resolution mode for the 2k$\times$2k Tektronix CCD
detector of FORS1 has a pixel size of 24 $\mu$m corresponding to
0.2\arcsec on the sky.

With both the EFOSC2 and FORS1 instruments, linear polarimetry is
performed by inserting in the parallel beam a Wollaston prism which
splits the incoming light rays into two orthogonally polarized
beams. Each object in the field has therefore two orthogonally
polarized images on the CCD detector, separated by 20$\arcsec$ for
EFOSC2 and 22$\arcsec$ for FORS1. To avoid image overlapping, one puts
at the telescope focal plane a special mask made of alternating
transparent and opaque parallel strips whose width corresponds to the
splitting. The object is positioned at the center of a transparent
strip which is imaged on a region of the CCD free of defects. The
final CCD image then consists of alternate orthogonally polarized
strips of the sky, two of them containing the polarized images of the
object itself (di Serego Alighieri {\cite{SER89}}, {\cite{SER97}};
Lamy \& Hutsem\'ekers {\cite{Lam99}}, hereafter Lam99). Note that the
polarization measurements do not depend on variable transparency or
seeing since the two orthogonally polarized images of the object are
simultaneously recorded.

In order to derive the two normalized Stokes parameters $q$ and $u$
which characterize the linear polarization, frames must be obtained
with at least two different orientations of the Wollaston. In
practice, the Wollaston is not rotated but a half-wave plate (HWP) is
inserted in the optical path and four frames are obtained with the HWP
at 4 different position angles (0$\degr$, 22.5$\degr$, 45$\degr$, and
67.5$\degr$). Even if only two different orientations of the HWP are
sufficient to retrieve the linear polarization (Melnick et al.
{\cite{MEL89}}), two additional orientations make it possible to
remove most of the instrumental polarization (di~Serego Alighieri
{\cite{SER89}}).

Targets were selected from the V\'eron catalogue (V\'eron-Cetty \&
V\'eron \cite{VER01}), from the Sloan Digital Sky Survey Early Data
Release (Schneider et al. \cite{SCH02,SCH03}, Reichard et
al. \cite{REI03}), from Becker et al. (\cite{BEC00}, \cite{BEC02}),
Menou et al. (\cite{MEN01}), Barkhouse \& Hall (\cite{BAR01}), Hall et
al. (\cite{HAL02}) and Smith et al. (\cite{SMI02}), mostly according
to their position on the sky i.e. with right ascensions and
declinations corresponding to the region of polarization vector
alignments defined in Hutsem\'ekers (\cite{HUT98a}). Bright objects
were preferred, as well as BAL, radio-loud and red QSOs which are
usually more polarized.

All but two observations were obtained through the Bessel V filter
with typical exposure times between 1 and 10 minutes per frame.
Polarized and unpolarized standard stars were observed in the Bessel
V, R, and gunn $i$ filters in order to unambiguously fix the
zero-point of the polarization position angle and to check the whole
observing and reduction process. In August 2000 and October 2003 the
sky was clear and the seeing around 1$\farcs$0, while in August 2003
conditions were not as good with cirrus and seeing around
1$\farcs$5. During the other runs, the seeing was always between
1$\farcs$1 and 1$\farcs$5, and the sky covered at worst with thin
cirrus. A few observations were obtained with a high Moon fraction
illumination ($>0.7$) and are of lower quality. Indeed, high levels of
sky background induce larger errors in the sky subtraction process and
the subsequent polarization measurements. This is especially relevant
when the polarization of the target is low.

\section {Data reduction}
\label{sec:data}

The $q$ and $u$ normalized Stokes parameters are computed from the
measurement of the integrated intensity of the orthogonally polarized
upper and lower images of the object, for the 4 different orientations
of the HWP. They are calculated with respect to the instrumental
reference frame using the following formulae:
\begin{eqnarray}
q & = & \frac{R_q - 1}{R_q + 1}  \hspace{0.5cm} \mbox{where} \hspace{0.5cm} 
R_q^2  = \frac{I_{\srm 0}^{\srm u}/I_{\srm 0}^{\srm l}}
       {I_{\srm 45}^{\srm u}/I_{\srm 45}^{\srm l}},\nonumber\\
 & &\\
u & = & \frac{R_u - 1}{R_u + 1}  \hspace{0.5cm} \mbox{where} \hspace{0.5cm} 
R_u^2  = \frac{I_{\srm 22.5}^{\srm u}/I_{\srm 22.5}^{\srm l}}
       {I_{\srm 67.5}^{\srm u}/I_{\srm 67.5}^{\srm l}},\nonumber 
\end{eqnarray}
where $I^{\srm u}$ and $I^{\srm l}$ respectively refer to the
intensities integrated over the upper and lower orthogonally polarized
images of the object. The combination of four frames obtained with
different HWP orientations not only removes most of the instrumental
polarization, but is also essential for correcting the effects of
image distortions introduced by the HWP (Lam99).

In order to measure levels of polarization as small as 0.6~\% with
0.2~\% uncertainty, it is mandatory to achieve photometry with a very
high accuracy. For this purpose, the data were first corrected for
bias and flat-fielded. The photometric measurements for each image
were done using the MIDAS procedure developed by Lam99. The different
steps of this procedure are the following : (1) several regions of the
background close to the target are interactively chosen; a plane is
fitted to their mean values and subtracted from each image
individually; (2)~the position of the object in each strip is measured
at subpixel precision by fitting a 2D gaussian profile. The flux is
subsequently integrated in circles centered at the fitted positions;
(3) the Stokes parameters are then computed for any reasonable value
of the aperture radius. Since they are found to be stable against radius
variation, it was decided to always measure them inside a fixed
aperture radius of $3.0 \times [(2\,\ln 2)^{-1/2}$ HWHM]
{\footnote{For He~1304-1157, the radius was $2.5 \times [(2\,\ln 2)
^{-1/2}$ HWHM] due to the presence of a cosmic-ray hit at larger radii.} 
where HWHM represents the mean half-width at half-maximum of the
gaussian profile. This empirical choice, seeing independent, is
unsensitive to image distorsions (Lam99).

The uncertainties $\sigma_q$ and $\sigma_u$ on the normalized Stokes
$q$ and $u$ are evaluated by computing the errors on the intensities
$I^{\srm u}$ and $I^{\srm l}$ from the read-out noise and from the
photon noise in the object and the sky background (after converting
the counts in electrons), and by propagating these
errors. Uncertainties are around 0.15\% for both $q$ and $u$.  For a
few faint objects listed in Table~{\ref{tab:rej}} we were not able to
derive reliable measurements, namely due to a higher than usual sky
background. These measurements are rejected from
the sample presented in Table~4.

A zero-point angle offset correction, filter dependent, is then
applied to the QSO normalized Stokes parameters $q$ and $u$ in order
to convert the polarization angle measured in the instrumental
reference frame into the equatorial reference frame. This angle offset
is determined from polarized standard stars observed each night and
listed in Table~{\ref{tab:std}}. These stars have been selected to
have polarization angles distributed in the full [0\degr, 180\degr]
range. For all stars observed during a given run (and between the runs
themselves), the values of the angle offset do agree within 1$\degr$
standard deviation \footnote{Except the measurements for HD251204
which disagree in both polarization degree and position angle from
tabulated data, possibly indicating polarization variability (see also
Weitenbeck \cite{WEI99})}.

We have also observed several unpolarized standard stars in the V, R, and
$i$ filters (Table~\ref{tab:std}). For these stars we measure $p =
0.12\pm0.05$\%, $p = 0.08\pm0.04$\% and $p = 0.11\pm0.04$\% in August
2000, August 2003 and October 2003, respectively, indicating that the
residual instrumental polarization is small, a result in agreement
with the expectation that most of the instrumental polarization is
removed by the observing procedure. No difference between the three
filters has been noticed.

\begin{table}[t]
\caption[ ]{The objects for which no reliable polarization 
measurements 
could be derived}
\label{tab:rej}
\begin{tabular}{ll}
\hline \\[-0.10in]
Date & Object \\ dd-mm-yyyy &
\\[0.05in] \hline \\[-0.10in]
27-08-2000 & \object{PKS 2357-318} 	\\
23-03-2002 & \object{SDSS J0948+0024}, \object{SDSS J1217-0029} 	\\
23-03-2002 & \object{SDSS J1235-0036}, \object{PKS 1308+145} 	\\
21-04-2002 & \object{SDSS J1217-0029}, \object{SDSS J1235-0036}	\\
02-05-2002 & \object{CTS A09.82} 		\\
20-08-2003 & \object{PKS 2357-318}       \\
21-08-2003 & \object{1WGA J2201.6-5646}  \\
22-08-2003 & \object{PKS 2314-116}       \\
19-10-2003 & \object{4C 19.74}           \\
\hline
\end{tabular}
\end{table} 

\begin{table}[tbp]
\caption[ ]{The observed standard stars}
\label{tab:std}
\begin{tabular}{llll}
\hline \\[-0.10in]
Date & Polarized & Unpolarized & Ref\\ dd-mm-yyyy & & &
\\[0.05in] \hline \\[-0.10in]

27-08-2000 & \object{HD155197}, \object{HD161056},         & \object{HD154892}  & 1 \\
           & \object{HD251204}, \object{BD25\degr+727}     &           & 1 \\
28-08-2000 & \object{HD155197}, \object{HD161056}          & \object{HD14069}   & 1 \\
20-03-2002 & \object{HD126593}, \object{HD298383} & & 1 \\
21-03-2002 & \object{HD111579}, \object{HD298383} & & 1 \\
22-03-2002 & \object{HD164740}, \object{HD298383} & & 1, 2 \\
23-03-2002 & \object{HD111579}, \object{HD155197} & & 1 \\
	   & \object{HD298383} 		& & 1 \\
21-04-2002 & \object{HD111579}, \object{HD155197} & & 1 \\
01-05-2002 & \object{HD155197}, \object{HD298383} & & 1 \\
02-05-2002 & \object{HD111579}, \object{HD298383} & & 1 \\
25-03-2003 & \object{Hiltner 652} & & 3 \\
20-08-2003 & \object{HD155197}, \object{HD251204}          & \object{HD154892}  & 1\\
21-08-2003 & \object{HD155197}, \object{HD126593}          &           & 1\\
22-08-2003 & \object{HD155197}, \object{BD25\degr+727}     & \object{HD154892}  & 1\\
19-10-2003 & \object{HD155197}, \object{BD25\degr+727}     & \object{HD64299}   & 1\\
19-10-2003 & \object{HD155197}, \object{HD298283}          & \object{HD64299}   & 1\\
\hline
\end{tabular}
{\footnotesize $ $ \\[0.1cm] References: (1) Turnshek et
al. {\cite{TUR90}}, (2) Serkowski et al. {\cite{SER75}}, (3)~Heiles
{\cite{HEI00}}}
\end{table} 

\begin{table}[tbp]
\caption[ ]{The residual polarization}
\label{tab:stars}
\begin{tabular}{lrrrr}
\hline \\[-0.10in]
Observing run & $\overline{q}_{\star}$ \ & $\overline{u}_{\star}$ \ & 
    $\overline{\sigma}_{\star}$ \ & $n_{\star}$ \\
 mm-yyyy    &     (\%) & (\%) & (\%) &
\\[0.05in] \hline \\[-0.10in]
08-2000 & $+$0.04 & $-$0.01 & 0.19 & 30 \\
03-2002 & $-$0.07 & $+$0.07 & 0.12 & 57 \\
05-2002 & $-$0.05 & $+$0.05 & 0.12 & 31 \\
08-2003 & $-$0.10 & $-$0.18 & 0.24 & 39 \\
10-2003 & $+$0.04 & $-$0.08 & 0.20 & 31 \\
\hline
\end{tabular}
\end{table} 

Since on most CCD frames field stars are simultaneously recorded, one
can use them to estimate the residual instrumental polarization and/or
interstellar polarization. While a frame-by-frame correction of the
QSO Stokes parameters is in principle possible, it is nevertheless
hazardous since we are never sure that the polarization of field stars
correctly represents the interstellar polarization which could affect
distant QSOs. For example, we have found stars located in the same
field with significantly different polarization degrees and
angles. Also, the field stars are often fainter than the QSO such that
a frame-by-frame correction would introduce uncertainties on the QSO
polarization larger than the correction itself. We then compute the
weighted average ($\overline{q}_{\star}$ and $\overline{u}_{\star}$)
and dispersion ($\overline{\sigma}_{\star}$) of the normalized Stokes
parameters of field stars considering the $n_{\star}$ frames with
suitable stars obtained during a given run. These quantities are given
in Table~{\ref{tab:stars}}.  The small values and dispersions of the
residual polarization confirm the small level of uncorrected
instrumental polarization. They also indicate that, on average, the
interstellar polarization is small, in agreement with the fact that
all objects in the sample are at high galactic
latitudes\footnote{Except FIRST~J0809+2753 with $|\bgal| =
28.33\degr$} ($|\bgal| \geq 30\degr$). To minimize the systematic
errors, this residual polarization is conservatively taken into
account by subtracting the systematic $\overline{q}_{\star}$ and
$\overline{u}_{\star}$ from the measured QSO $q$ and $u$, and by
adding quadratically the errors. Since $\overline{q}_{\star}$ and
$\overline{u}_{\star}$ are nearly identical in March and May 2002,
only the mean values $\overline{q}_{\star} = -0.06$\%,
$\overline{u}_{\star} = +0.06$\% and $\overline{\sigma}_{\star} =
0.12$\% are used.  For those objects observed with FORS1 in February
2003, no correction is done, while the correction used for the March
and May 2002 runs is applied to the April 2002 data, obtained with the
same instrumental setup.

\begin{figure*}[t]
\begin{center}
{\includegraphics*[width=14cm]{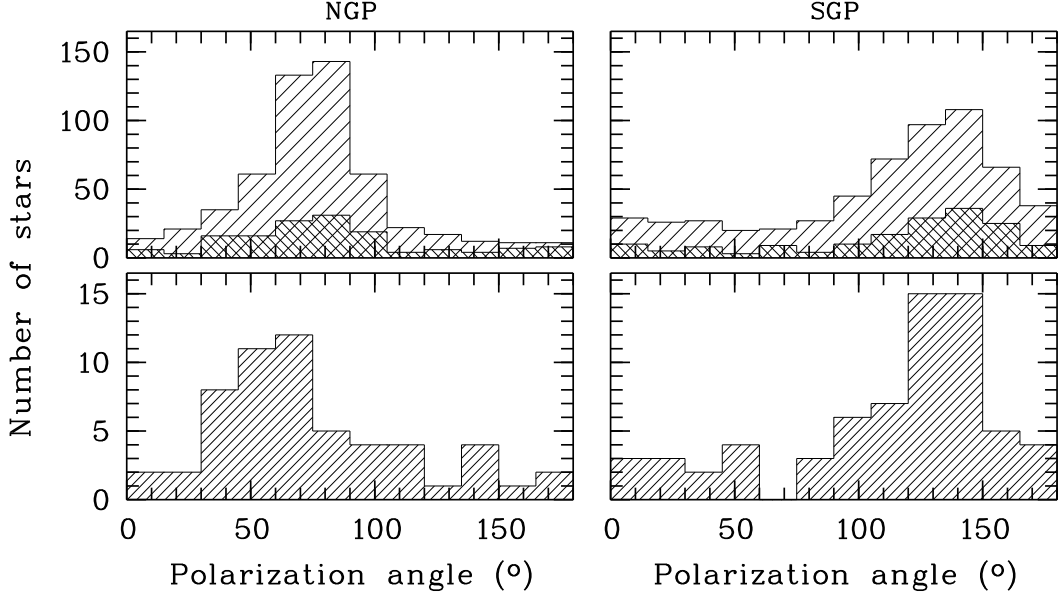}}
\end{center}
\caption{$ $ {\it Top:} the distribution of the polarization angles of
stars from the Heiles catalogue (Heiles \cite{HEI00}). Only high
latitude stars ($|\bgal| \geq 30^{\circ}$) with polarization angles
having uncertainties $\sigma_{\theta} \leq 14^{\circ}$ are
considered. The NGP and SGP regions are illustrated separately. The
darker histograms refer to stars at distances $\geq 200$~pc.  {\it
Bottom:} the distribution of the polarization angles of field stars
measured on QSO frames. Data from previous papers are included. Only
those objects with uncertainties on polarization angles
$\sigma_{\theta} \leq 14^{\circ}$ are illustrated.}
\label{fig:histar}
\end{figure*}

In order to better understand the nature and the effect of this
correction, we illustrate in Fig.~\ref{fig:histar} (bottom panel) the
distribution of the polarization position angles we have measured for
field stars\footnote{We consider a single star per frame/field. In
some cases this star is made up of the combination of several fainter
stars from a given frame.} at high galactic latitude ($| \bgal | \ge
30^{\circ}$).  Polarization data reported in Lamy \& Hutsem\'ekers
(\cite{Lam00}, hereafter Lam00) and obtained with the same
instrumentation are also included. After removing bad quality
measurements and redundancies, this leads to a total sample of 204
field star measurements at $| \bgal | \ge 30^{\circ}$, of which about
half have polarization angles with $\sigma_{\theta} \leq 14^{\circ}$.
The polarization angles of the stars from the Heiles catalogue (Heiles
\cite{HEI00}) are illustrated in the top panel of
Fig.~\ref{fig:histar}. The distributions are very similar to
ours\footnote{There is also a good agreement between the polarization
degrees (typically around 0.2--0.3\% , cf. Fig.~\ref{fig:pstar}),
provided that one considers distant stars in the Heiles catalogue,
i.e. stars at distances $\geq 100-200$~pc.}, including definite
concentrations of the polarization angles around two main directions:
$\sim$70$^{\circ}$ towards the NGP and $\sim$135$^{\circ}$ towards the
SGP.  The existence of these two major directions in the interstellar
polarization towards the North and the South Galactic Poles has also
been reported by Berdyugin et al. (\cite{BER04}), considering distant
stars at high galactic latitudes.  This similarity in the polarization
angle distributions suggests that a significant part of the
polarization we measure for field stars is interstellar in origin.  If
we average the values of the residual polarization for the NGP and the
SGP separately, we get from Table~\ref{tab:stars} and Lam00,
$\overline{q}_{\star}$ = $-0.05$\% and $\overline{u}_{\star}$ =
$+0.08$\% for the NGP, and $\overline{q}_{\star}$ = $-0.01$\% and
$\overline{u}_{\star}$ = $-0.09$\% for the SGP, which correspond to
the polarization angles $\overline{\theta}_{\star}$ = 61$\degr$ and
$\overline{\theta}_{\star}$ = 133$\degr$, respectively, in agreement
with the trend seen in Fig.~\ref{fig:histar}.  Note that a small
contribution due to instrumental polarization cannot be excluded given
the differences in the values of $\overline{q}_{\star}$ and
$\overline{u}_{\star}$ for the various runs (Table~\ref{tab:stars},
Lam00).  Although small, the correction by the systematic
$\overline{q}_{\star}$ and $\overline{u}_{\star}$\footnote{Different
$\overline{q}_{\star}$ and $\overline{u}_{\star}$ are used for each
observing run. Ideally, as suggested by the results in
Fig.~\ref{fig:histar}, one should have also computed
$\overline{q}_{\star}$ and $\overline{u}_{\star}$ for $\bgal \ge
30^{\circ}$ and $\bgal \le -30^{\circ}$ separately. However this makes
little difference since nearly all objects observed in a given run
were either at $\bgal \ge 30^{\circ}$ or at $\bgal \le -30^{\circ}$.}
then removes the bias in the distribution of the polarization angles
observed in Fig.~\ref{fig:histar}, at least from the statistical,
systematic, point of view.

\section {The results}

\input{2163_4sissatab}

\begin{figure}[t]
\resizebox{\hsize}{!}{\includegraphics*{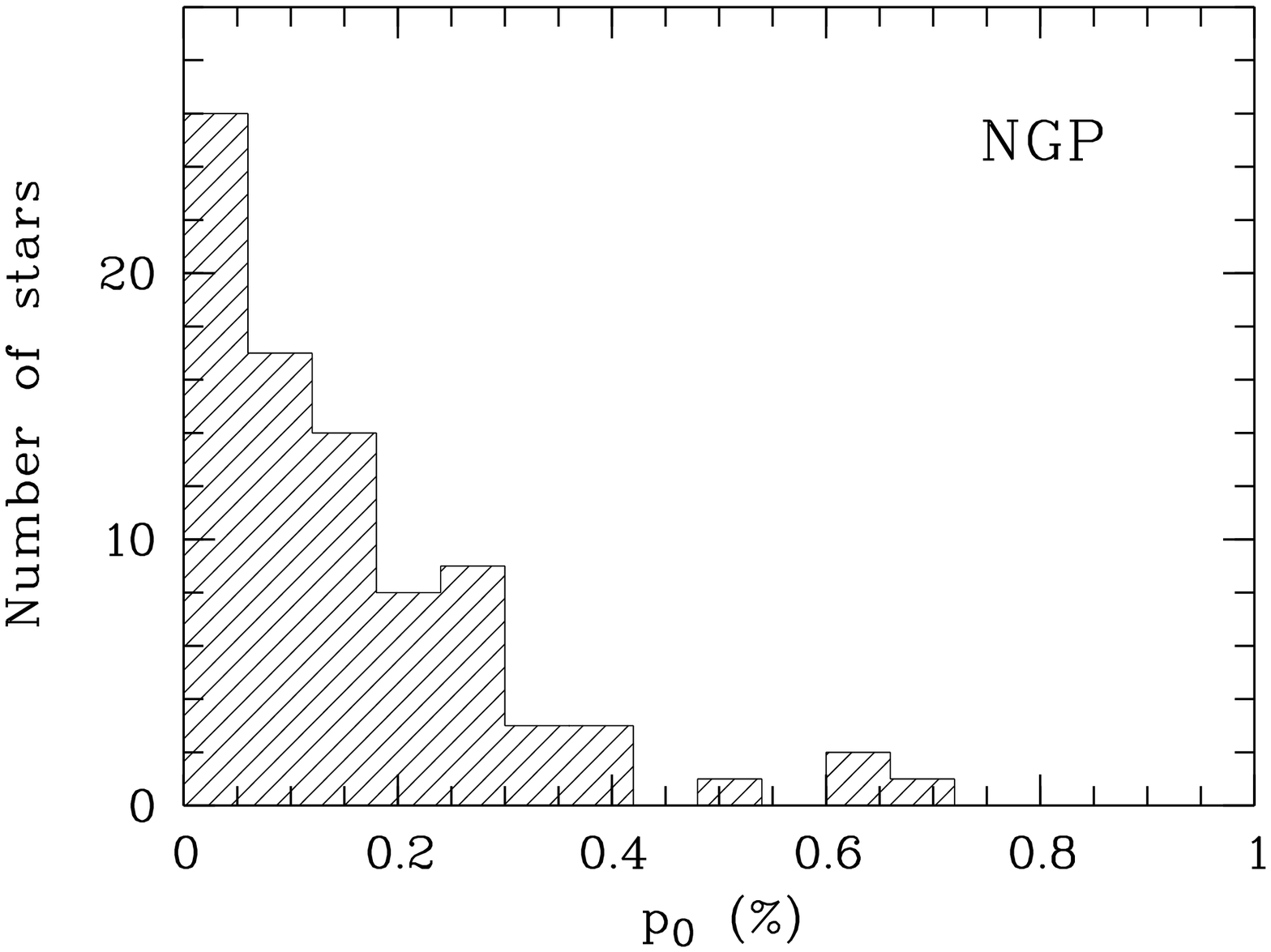}} \\[0.5cm]
\resizebox{\hsize}{!}{\includegraphics*{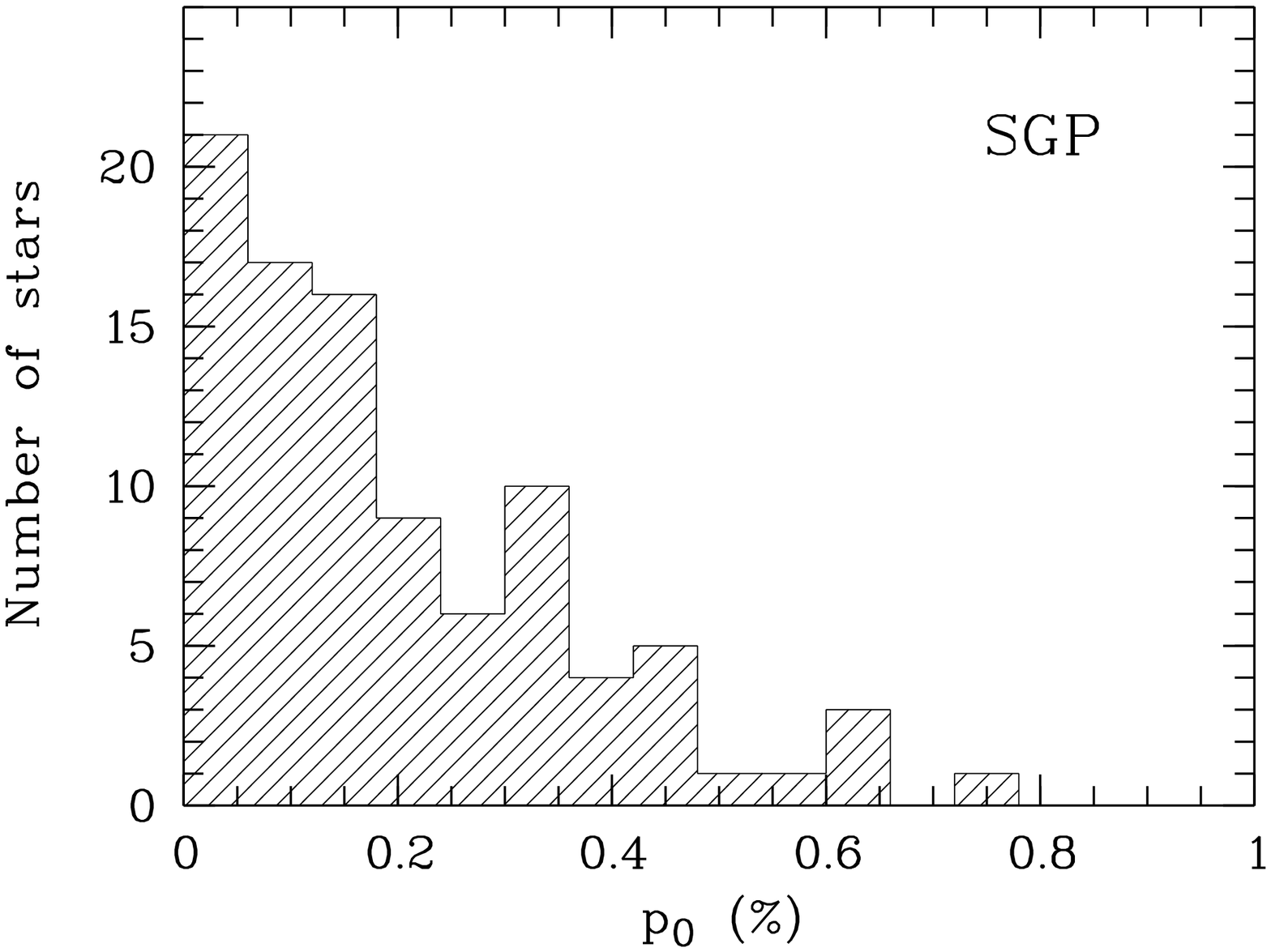}}
\caption{The distribution of the debiased polarization degree of field
stars in the NGP region (top, from the 2002 data) and in the SGP
region (bottom, from the 2000 \& 2003 data).  The correction by the
systematic $\overline{q}_{\star}$ and $\overline{u}_{\star}$ is
done. Only measurements with an error on the polarization degree $<$
0.3\% are represented.}
\label{fig:pstar}
\end{figure}

Table~4 summarizes the measurements. The first eight
columns give the QSO name, type, equatorial coordinates (J2000) and
redshift $z$, the date of observation and the normalized Stokes
parameters $q$ and $u$ corrected for the systematic residual
polarization given in Table~\ref{tab:stars}. The normalized Stokes
parameters are given in the equatorial reference frame. The QSO name
is the one used in the V\'eron catalogue (V\'eron-Cetty and V\'eron
{\cite{VER01}}) when given, and in the NASA/IPAC Extragalactic
Database (NED) otherwise. The name is followed by the object
classification.  The following notation has been adopted : {\it R} if
known radio emitter, {\it B} if known BAL, {\it RB} if both, and {\it
U} otherwise.

Then, from these values, the polarization degree is evaluated with $p
= (q^2+u^2)^{1/2}$ and the associated error with $\sigma_p = (\sigma^2
+ \overline{\sigma}^2_{\star})^{1/2}$ where $\sigma \simeq \sigma_q
\simeq \sigma_u$.  In addition, $p$ must be corrected for the
statistical bias inherent to the fact that $p$ is always a positive
quantity. For this purpose, we used the Wardle \& Kronberg estimator
({\cite{WAR74}}) which was found to be a reasonably good estimator of
the true polarization degree (Simmons \& Stewart {\cite{SIM85}}). The
debiased value $p_{0}$ of the polarization degree is reported in
column 11. The polarization position angle $\theta$ is obtained by
solving the equations $q = p\cos 2\theta$ and $u = p \sin
2\theta$. The uncertainty of the polarization position angle $\theta$
is estimated from the standard Serkowski ({\cite{SER62}}) formula
where the debiased value $p_{0}$ is conservatively used instead of
$p$, i.e. $\sigma_{\theta} = 28\fdg65\, \sigma_p / p_{0}$. Note that
due to the HWP chromatism over the V band, an additional error
$\leqslant 2-3\degr$ on $\theta$ should be accounted for (cf. the
wavelength dependence of the polarization angle offset in di Serego
Alighieri {\cite{SER97}}).

Fig.~\ref{fig:pstar} illustrates the distribution of the field star
polarization measured on the QSO frames\footnote{We were also able to
measure the polarization of 10 field galaxies with a reasonable
accuracy. Within the uncertainties, their polarization (both in degree
and angle) does not differ from the polarization of the field
stars. This suggests that the contamination by interstellar
polarization in our Galaxy is not significantly higher for objects at
extragalactic distances.}, for the NGP and the SGP regions
separately. For both regions of the sky, the polarization degree is
small, most often $\leqslant$ 0.3\%.  The distributions are very
similar although stars with polarization between 0.3\% and 0.5\% seem
slightly more numerous in the SGP. The median polarization is 0.11\%
in the NGP region and 0.15\% in the SGP region. The overall
distribution suggests that virtually every quasi-stellar object with a
polarization higher than 0.6\% is intrinsically polarized, in
agreement with previous studies (Berriman et al. \cite{BER90},
Lam00). In only five cases (flagged in Table~4), the QSO
polarization is both significant ($ \geqslant 0.6$~\%) and comparable
in degree and angle to the polarization of field stars, suggesting a
probable contamination. Four objects with marginal contamination are
also indicated in that Table.

\subsection{Testing for possible biases in the data}

Since the goal of our polarization measurements is to study
concentrations of QSO polarization position angles along preferred
directions, it is important to verify that the distribution of
polarization angles is not significantly contaminated by either the
instrumental polarization or the interstellar polarization in our
Galaxy.

\subsubsection{Instrumental polarization}

\begin{figure}[t]
\centering
\resizebox{\linewidth}{!}{\includegraphics*{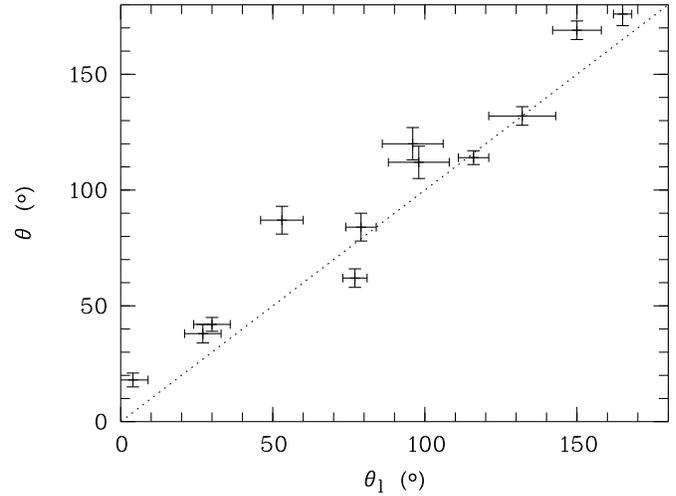}}
\caption{ The measured polarization angle $\theta$ (from
Table~4) is reported against the published value
$\theta_l$ (from Table~{\ref{tab:conf}}). Due to its very low
polarization level and highly uncertain $\theta$ in
Table~4, \object{2MA~J1714+2602} is not represented here.}
\label{fig:conf}
\end{figure}

\addtocounter{table}{+1}
\begin{table}[tbp]
\caption[ ]{Published ``white light'' 
polarization data for the verification objects}
\label{tab:conf}
\begin{tabular}{lllrrl}
\hline \\[-0.10in]
Object & $p_l$ & $\sigma_{p_l}$ & $\theta_l$ & $\sigma_{\theta_l}$ & Ref\\
       & (\%)  &  (\%)          & (\degr)    &  (\degr) &
\\[0.05in] \hline \\[-0.10in]
\object{PG 0946+301} & 0.85 & 0.14 & 116 & 5 & 1,2\\
\object{PG 1004+130} & 0.79 & 0.11 & 77 & 4 & 2\\
\object{PG 1012+008} & 0.66 & 0.23 & 98 & 10 & 2\\
\object{PKS 1049-09} & 0.85 & 0.30 & 96 & 10 & 2\\
\object{PG 1216+069} & 0.80 & 0.19 & 53 & 7 & 2\\
\object{TON 1530} & 0.84 & 0.24 & 150 & 8& 2 \\
\object{PG 1254+047} & 1.22 & 0.15 & 165 & 3& 2,1\\
\object{PG 1435-067} & 1.44 & 0.29 & 27 & 6& 2\\
\object{2MA 1519+1838}& 0.67 & 0.22 & 132 & 11& 3\\
\object{2MA 1543+1937}& 1.33 & 0.26 & 30 & 6 &3\\
\object{3C 323.1} & 1.03 & 0.20 & 4 & 5 &2 \\
\object{MARK 877} & 0.94 & 0.17 & 79 & 5 &2 \\
\object{2MA J1714+2602}& 0.86 & 0.33 & 65 & 12 &3 \\ 
\hline
\end{tabular}
{\footnotesize $ $ \\[0.1cm] References: (1) Schmidt \& Hines
{\cite{SCH99}}; (2) Berriman et al. {\cite{BER90}}; (3)~Smith et
al. {\cite{SMI02}}.  When more than one measurement is available
(cf. \object{PG 0946+301} and \object{PG 1254+047}), the value with the smallest
$\sigma_{p_l}$ is given.}
\end{table} 

To verify that the polarization angles are not affected by an
instrumental bias, a first test was done using standard stars with
intrinsic polarization angles distributed in the full [0\degr,
180\degr] range (Sect.~\ref{tab:std} and Lam00). The excellent
agreement --most often within 1$\degr$-- between the polarization
angles we measure and those values published in the literature
demonstrates the absence of such a bias, at least for highly polarized
objects.

\begin{figure*}[!t]
\resizebox{0.9\linewidth}{!}{\includegraphics*{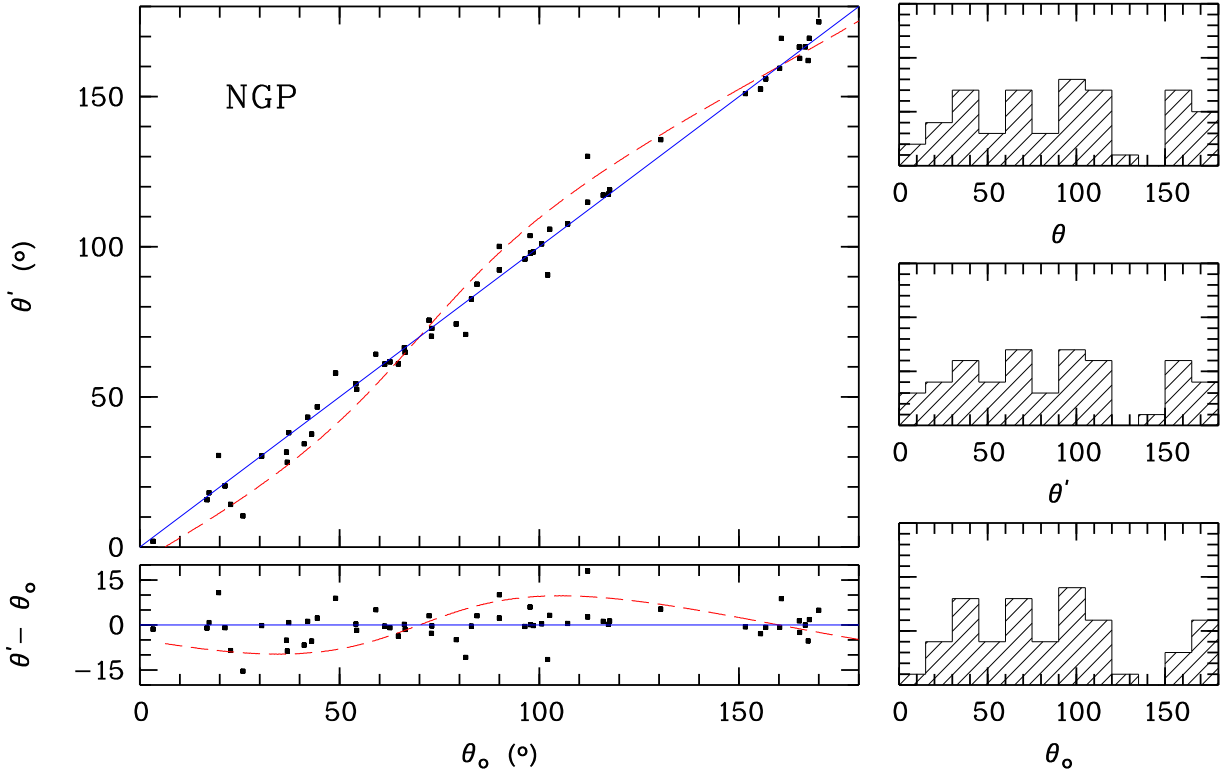}}
\caption{ \ This figure compares, for the NGP region (2002 data), the
QSO polarization angle $\theta_{0}$, uncorrected for the systematic
$\overline{q}_{\star}$ and $\overline{u}_{\star}$, to the polarization
angle $\theta '$ derived after a frame by frame subtraction of the
Stokes parameters of field stars (when the latter measurements were
possible). Only QSOs with either $p_0 \geq 0.6$\% or $p' \geq 0.6$\%
are plotted.  For most objects, $|\theta ' - \theta_0 |$ is
within~5$^{\circ}$. As a reference, the dashed line represents
simulated data obtained by subtracting a systematic polarization of
0.2\% oriented at 70\degr\ (cf. Fig.~\ref{fig:histar}) from a
polarization of 0.6\% randomly oriented.  Together with the
distributions of $\theta_{0}$ and $\theta '$, the histograms
illustrate the distribution of polarization angle $\theta$ reported in
Table~4 and corrected for the systematic
$\overline{q}_{\star}$ and $\overline{u}_{\star}$ (QSOs with $p \geq
0.6$\% only).}
\label{fig:testn}
\end{figure*}

\begin{figure*}[!t]
\resizebox{0.9\linewidth}{!}{\includegraphics*{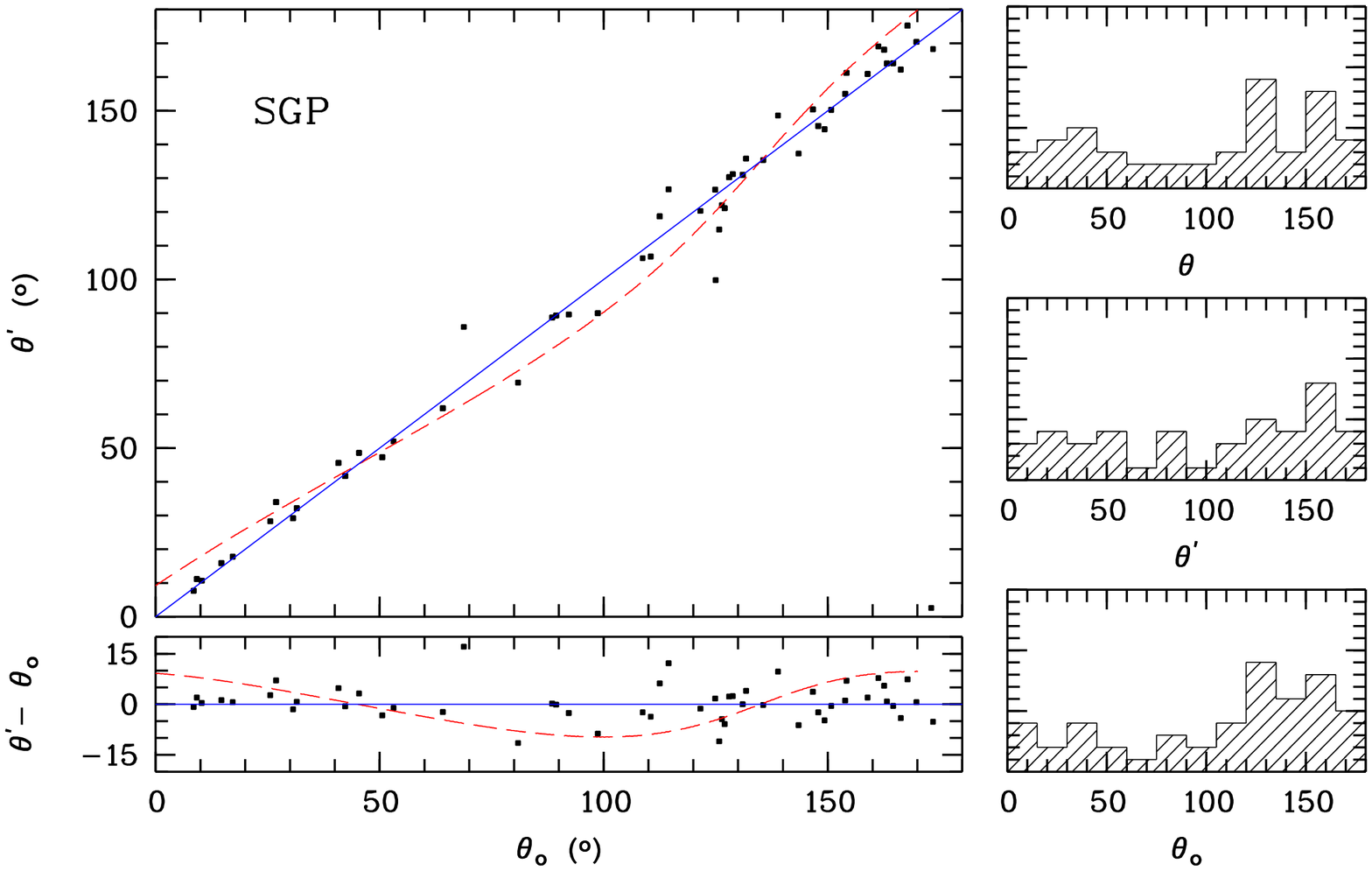}}
\caption{ \ Same as Fig.~\ref{fig:testn}, but for QSOs from the SGP
region (2000 \& 2003 data). For most objects, $|\theta ' - \theta_0 |$
is also within~5$^{\circ}$. The dashed line represents simulated data
obtained by subtracting a systematic polarization of 0.2\% oriented at
135\degr\ (cf. Fig.~\ref{fig:histar}) from a polarization of 0.6\%
randomly oriented. }
\label{fig:tests}
\end{figure*}

In order to perform a similar test at lower polarization levels,
typically around 1\%, we have measured the polarization of a sample of
13 QSOs previously observed by Berriman et al. ({\cite{BER90}}),
Schmidt \& Hines ({\cite{SCH99}}) and Smith et al. ({\cite{SMI02}})
using different telescopes and instruments. The targets were selected
to have polarization angles distributed in the full [0\degr, 180\degr]
range (Table~{\ref{tab:conf}}).  Fig.~{\ref{fig:conf}} shows the
observed polarization angle $\theta$ (from Table~4) versus the
published one $\theta_l$ (from Table~{\ref{tab:conf}}).  A good
overall agreement is observed, suggesting the absence of an
instrumental bias.  However, the correlation is not as good as one
could have expected given the formal uncertainties. This is due to the
fact that we compare unfiltered, white light, measurements to V-band
polarization angles (with some additional noise from possible time
variability).  Indeed, while basically constant with wavelength, a
slight dependence of the polarization angle on wavelength (i.e. about
ten degrees between the blue and red parts of the spectrum) is often
observed in low polarization QSOs (Webb et al. {\cite{WEB93}},
Antonucci et al. \cite{ANT96}).  For two objects, V-band polarization
angle measurements are available in the literature, $\theta$ = 63
$\pm$ 3\degr\ for \object{PG~1004+130} (Webb et al. {\cite{WEB93}})
and $\theta$ = 23 $\pm$ 2\degr\ for \object{3C~323.1} (Schmidt \&
Smith \cite{SCH00}). They are in much better agreement with our V-band
measurements than with the white light data.  A comparison of
polarization measurements obtained in the same filter is then needed
to test more accurately for possible instrumental contamination at low
polarization levels.

\subsubsection{Interstellar polarization}

In order to estimate the effect of the interstellar polarization on
significantly polarized ($p \geq 0.6$\%) objects, and namely whether
it can introduce a bias in the distribution of the QSO polarization
angles, we subtract frame by frame (when possible) the Stokes
parameters of the field stars from the QSO ones, both uncorrected for
the systematic $\overline{q}_{\star}$ and $\overline{u}_{\star}$.  New
QSO polarization degrees ($p'$) and angles ($\theta '$) are then
derived. With such a test, we implicitely assume that the polarization
of field stars correctly represents the interstellar polarization
affecting the QSOs.

In Fig.~\ref{fig:testn}~\&~\ref{fig:tests} we compare the new
polarization angle $\theta '$ to the uncorrected one $\theta_0$, for
the NGP and the SGP regions separately.  Only significantly polarized
QSOs are considered, i.e. those objects with a polarization degree
$\geq 0.6\%$.  The distribution of the final polarization angle
$\theta$ given in Table~4 is also illustrated.  Although a
systematic effect may be noticed, it is small enough to only slightly
modify the distribution of the QSO polarization angles. This is due to
the fact that the field star polarization is most often small
(Fig.~\ref{fig:pstar}) such that only a few significantly polarized
QSOs are affected (and, as done in Table~4, the most
discordant objects may be flagged as contaminated).

This test demonstrates that the distribution of the polarization
angles of polarized ($p \geq 0.6$\%) QSOs is not significantly
contaminated by the interstellar polarization, and more particularly
after subtracting the systematic correction. But this conclusion is
only valid if the interstellar polarization which affects the QSOs is
not significantly larger than the polarization measured from field
stars.

\subsection{Time variability of the polarization}

For some targets of our sample, several polarimetric measurements do
exist.  \object{MARK~877}, the BAL QSO \object{Q2208-1720}, and the binary QSO
\object{PKS~1145-071} A\&B have been observed at two epochs (March and May
2002) and do not show any evidence for significant polarization
variability. Also, the measurements for \object{MARK~877} and the BAL QSO
\object{J2359-12} are in excellent agreement with the values reported by
Berriman et al. (\cite{BER90}) and Brotherton et al. (\cite{BRO01}),
respectively.  The four radio-emitters \object{PKS 2203-215},
\object{PKS 2204-54}, \object{PKS 2240-260} and \object{PKS~1136-13}
were previously observed in white light by Fugmann \& Meisenheimer
(\cite{FUG88}) and Impey \& Tapia (\cite{IMP90}). Within the
uncertainties and given the different wavelength coverages, we find a
good agreement with our polarimetric measurements, i.e.  no evidence
for a significant variability, including for the highly polarized
quasar \object{PKS 2240-260}.

On the contrary, the polarization level of \object{PKS~1222+037} grew
up to 2.4~\% between March and May 2002, indicating that this object
might be variable. Our observations also suggest the variability (in
both polarization level and angle) of the BL Lac candidate
\object{PKS~1216-010} (Londish et al. \cite{LON02}). Indeed,
Visvanathan \& Wills ({\cite{VIS98}}) report for this object a
polarization level $p = 6.9
\pm 0.8$\% and a polarization angle $\theta = 8 \pm 3.3\degr$ which are
significantly different from our values ($p_0 = 11.1 \pm 0.19$\% and
$\theta = 100 \pm 0.5\degr$). Such a variability is common among
highly polarized quasars and BL Lac objects (see e.g. Rieke et
al. \cite{RIE77}, Valtaoja et al. \cite{VAL91}).

\section{Conclusions}

New polarization measurements have been obtained for a sample of 203
quasi-stellar objects located in the NGP and the SGP regions with a
final median uncertainty $\simeq$~0.25\% on the polarization
degree. 184 measurements are first time measurements. Among these
ones, half of the 42 BAL QSOs show a level of polarization $>$ 1\% and
12 QSOs have a polarization level higher than 3\% (including the BAL
QSO \object{SDSS~J1409+0048}).

Based on previous measurements found in the literature, we find
evidence for a variation of both the polarization level and the
polarization angle of the BL Lac candidate \object{PKS~1216-010}. We
also report a significant variability in a two month period of the
polarization of \object{PKS~1222+037}. Such a variability is not
surprising since this object is reported to have a flat radio spectrum
(Teraesranta et al. \cite{TER01}) which means, in unified schemes, a
pole-on view of the AGN with a relativistically beamed polarized
emission (Urry \& Padovani \cite{URR95}, Rusk \cite{RUS90}, Antonucci
\& Ulvestad \cite{ANT85}).

The pair of quasi-stellar objects \object{PKS~1145-071} A\&B has been
observed at two different epochs. They have polarization levels
$>0.6$\% and different polarization angles.

This sample also includes the gravitationally lensed QSO
\object{J11319-1231} (Sluse et al. \cite{SLU03}). The lensed nature of
this object has been serendipitously unveiled on the polarimetric
images obtained during the May 2002 observing run. The global
polarization level of this object is smaller than 0.3\%, confirming
that gravitationally lensed QSOs are not more polarized than other
quasi-stellar objects (cf. Hutsem\'ekers et al. \cite{HUT98b} and
Lam00).

We have also shown that, if the effect of the interstellar
polarization in our Galaxy onto distant objects may be adequately
represented by the polarization measured from field stars, the
significantly polarized ($p \geq 0.6$\%) QSOs show little
contamination in the distribution of their polarization angles.

\begin{acknowledgements}
Dominique Sluse acknowledges support from an ESO studentship in
Santiago and PRODEX (Gravitational lens studies with HST).
Herv\'e Lamy would like to thank Prof. J.~Lemaire and BIRA-IASB for
giving him the opportunity to observe in La Silla in March 2002.
Hernan Quintana acknowledges partial support from the FONDAP Centro 
de Astrophysics. 
This research has made use of the NASA/IPAC Extragalactic Database
(NED), which is operated by the Jet Propulsion Laboratory, California
Institute of Technology, under contract with the National Aeronautics
and Space Administration.

\end{acknowledgements}


\end{document}

%% file: 2163_4sissatab.tex
\begin{table*}[t]
\centering
\caption{The polarization measurements}
\label{tab:gen}
\begin{tabular}{llccccrrrrrrrr}
\hline 
Object  &  & RA J2000 & DEC J2000 & $z$ & Date & $q$ \ \ & $u$ \ \ & $p$ \ \ & ${\sigma}_p$  & $p_0$ \ & $\theta$ \  & ${\sigma}_{\theta}$ &    \\
& & (h m s)& $(\degr$ $\arcmin$ $\arcsec$) & & dd-mm-yyyy & (\%)  & (\%) & (\%) & (\%) & (\%) &  ($\degr$) & ($\degr$) &  \\ 
 \hline \\

PHL 850                & R &  00 49 32.1 & $+$11 28 26 & 0.275  &  28-08-2000 &     0.13 &  $-$0.41 &      0.43 &     0.23 &     0.38 &     144 &    17 & \\ 
RXS J01526-0143        & R &  01 52 37.1 & $-$01 43 58 & 0.850  &  19-10-2003 &     0.03 &  $-$0.13 &      0.13 &     0.21 &     0.00 &       - &     - & \\
3C 059                 & R &  02 07 02.2 & $+$29 30 46 & 0.110  &  28-08-2000 &  $-$0.64 &  $-$0.86 &      1.07 &     0.21 &     1.05 &     117 &     6 & \\ 
UM 416                 & R &  02 16 12.3 & $-$01 05 19 & 1.480  &  19-10-2003 &  $-$0.13 &  $-$0.12 &      0.17 &     0.25 &     0.00 &       - &     - & \\
J0235.2-0357           & R &  02 35 14.0 & $-$03 57 34 & 1.387  &  19-10-2003 &     0.08 &     0.89 &      0.90 &     0.23 &     0.87 &      42 &     8 & \\
SDSS J0242+0049        & B &  02 42 21.9 & $+$00 49 12 & 2.071  &  19-10-2003 &     1.33 &  $-$0.63 &      1.47 &     0.24 &     1.45 &     167 &     5 & \\
SDSS J0243+0000        & B &  02 43 04.7 & $+$00 00 05 & 2.003  &  20-10-2003 &     0.11 &     1.68 &      1.69 &     0.36 &     1.65 &      43 &     6 & \\
MS 02448+1928          & R &  02 47 40.8 & $+$19 40 58 & 0.176  &  27-08-2000 &  $-$0.01 &  $-$0.26 &      0.26 &     0.21 &     0.19 &     134 &    32 & \\ 
US 3498                & R &  03 00 29.8 & $+$02 40 50 & 0.115  &  27-08-2000 &  $-$0.02 &     0.55 &      0.55 &     0.27 &     0.49 &      46 &    16 & \\ 
FIRST03133+0036        & R &  03 13 18.6 & $+$00 36 23 & 1.250  &  20-10-2003 &  $-$0.50 &  $-$1.40 &      1.48 &     0.29 &     1.45 &     125 &     6 & \\
FIRST03145+0117        & R &  03 14 30.0 & $+$01 17 36 & 1.210  &  20-10-2003 &     0.44 &     0.26 &      0.51 &     0.36 &     0.41 &      16 &    25 & \\
SDSS J0318-0015        & B &  03 18 28.9 & $-$00 15 23 & 1.990  &  19-10-2003 &  $-$0.38 &     0.39 &      0.55 &     0.26 &     0.49 &      67 &    15 & \\
0321-375               & U &  03 23 53.4 & $-$37 15 57 & 2.246  &  28-08-2000 &     0.45 &     0.04 &      0.45 &     0.28 &     0.38 &       3 &    21 & \\ 
PKS 0446-212           & R &  04 48 17.4 & $-$21 09 45 & 1.971  &  20-10-2003 &     0.24 &     0.08 &      0.25 &     0.24 &     0.10 &      10 &    72 & \\
MC 0446-208            & U &  04 48 58.8 & $-$20 44 46 & 1.896  &  20-10-2003 &     0.61 &  $-$0.07 &      0.61 &     0.24 &     0.57 &     177 &    12 & \\ 
FIRST J0809+2753      & RB &  08 09 01.4   &  $+$27 53 41   &  1.511    & 21-03-2002   &$-$1.46  &   0.96   &   1.75   &  0.20  &   1.74   &   73 &    3 &\\
FIRST J0910+2612      & R  &  09 10 44.9   &  $+$26 12 53   &  2.920    & 01-05-2002   &$-$0.70  &$-$0.25   &   0.74   &  0.22  &   0.71   &  100 &    9 &\\
TEX 0907+258          & R  &  09 10 55.3   &  $+$25 39 21   &  2.743    & 01-05-2002     & 0.33  &$-$0.27   &   0.43   &  0.22  &   0.38   &  160 &   17 &\\
PG 0946+301           & B  &  09 49 41.1   &  $+$29 55 19   &  1.220    & 21-03-2002   &$-$1.12  &$-$1.21   &   1.65   &  0.19  &   1.64   &  114 &    3 &\\
FIRST J1000+2752      & R  &  10 00 29.1   &  $+$27 52 11   &  1.202    & 01-05-2002   &$-$0.21  &   0.09   &   0.23   &  0.20  &   0.15   &   78 &   38 &\\
FIRST J1003+2727      & R  &  10 03 18.9   &  $+$27 27 34   &  1.283    & 01-05-2002   &$-$0.61  &   0.44   &   0.75   &  0.20  &   0.73   &   72 &    8 &\\
PG 1004+130          & RB  &  10 07 26.2   &  $+$12 48 56   &  0.240    & 22-03-2002   &$-$0.51  &   0.76   &   0.92   &  0.13  &   0.91   &   62 &    4 &\\
PG 1012+008           & R  &  10 14 54.9   &  $+$00 33 37   &  0.185    & 22-03-2002   &$-$0.44  &$-$0.44   &   0.62   &  0.14  &   0.61   &  112 &    7 &\\
PKS 1012+022          & R  &  10 15 15.7   &  $+$01 58 52   &  1.374    & 01-05-2002   &$-$0.06  &$-$0.36   &   0.36   &  0.15  &   0.34   &  130 &   13 &\\
Q 1015+0147           & R  &  10 17 42.4   &  $+$01 32 17   &  1.455    & 01-05-2002     & 0.65  &$-$0.57   &   0.86   &  0.22  &   0.84   &  159 &    7 &$\star$\\
PKS 1049-09           & R  &  10 51 29.9   &  $-$09 18 09   &  0.345    & 22-03-2002   &$-$0.33  &$-$0.56   &   0.65   &  0.15  &   0.63   &  120 &    7 &\\
PKS 1115-12           & R  &  11 18 17.1   &  $-$12 32 54   &  1.739    & 20-03-2002   &$-$0.28  &   0.28   &   0.40   &  0.22  &   0.34   &   67 &   19 &\\
PKS 1118-05           & R  &  11 21 25.1   &  $-$05 53 56   &  1.297    & 21-03-2002   &$-$1.24  &$-$0.06   &   1.24   &  0.48  &   1.15   &   91 &   12 &\\
He 1122-1315          & U  &  11 25 09.4   &  $-$13 32 06   &  0.458    & 22-03-2002   &$-$1.21  &$-$0.92   &   1.52   &  0.15  &   1.51   &  109 &    3 &\\
SDSS J1126+0034       & B  &  11 26 02.8   &  $+$00 34 18   &  1.782    & 20-03-2002   &$-$0.23  &$-$0.41   &   0.47   &  0.21  &   0.43   &  120 &   14 &\\
PKS 1124-186          & R  &  11 27 04.4   &  $-$18 57 17   &  1.048    & 20-03-2002     & 3.15  &  11.25   &  11.68   &  0.36  &  11.68   &   37 &    1 &\\
R07.16                & U  &  11 28 18.5   &  $-$13 19 29   &  0.351    & 01-05-2002   &$-$0.48  &   0.33   &   0.58   &  0.15  &   0.56   &   73 &    8 &\\
PKS 1126+10           & R  &  11 29 14.2   &  $+$09 52 00   &  1.515    & 22-03-2002     & 0.25  &$-$0.38   &   0.45   &  0.20  &   0.42   &  152 &   14 &\\
He 1127-1304          & R  &  11 30 19.9   &  $-$13 20 51   &  0.634    & 01-05-2002   &$-$0.03  &   1.32   &   1.32   &  0.13  &   1.31   &   46 &    3 &\\
J11319-1231           & R  &  11 31 51.6   &  $-$12 31 57   &  0.654    & 02-05-2002   &$-$0.04  &   0.12   &   0.13   &  0.19  &   0.00   &   -  &   -  &\\
Q 1129-0229           & U  &  11 32 30.1   &  $-$02 46 21   &  0.333    & 02-05-2002     & 0.52  &   0.11   &   0.53   &  0.16  &   0.51   &    6 &    9 &\\
PKS 1131-17           & R  &  11 34 23.5   &  $-$17 27 51   &  1.618    & 21-03-2002     & 0.07  &   0.84   &   0.84   &  0.26  &   0.80   &   43 &    9 &\\
Q 1131-0039           & U  &  11 34 32.3   &  $-$00 55 49   &  0.268    & 01-05-2002   &$-$0.31  &$-$0.29   &   0.42   &  0.18  &   0.39   &  112 &   13 &\\
SDSS J1135+0041       & B  &  11 35 37.6   &  $+$00 41 30   &  1.550    & 20-03-2002     & 1.00  &   0.57   &   1.15   &  0.30  &   1.11   &   15 &    8 &\\
PKS 1134+01           & R  &  11 37 29.6   &  $+$01 16 14   &  0.430    & 02-05-2002     & 0.95  &$-$0.60   &   1.12   &  0.26  &   1.09   &  164 &    7 &\\
PKS 1136-13           & R  &  11 39 10.7   &  $-$13 50 43   &  0.554    & 23-03-2002     & 0.33  &   0.28   &   0.43   &  0.15  &   0.41   &   20 &   11 &\\
PKS 1145-071 A        & R  &  11 47 51.5   &  $-$07 24 41   &  1.342    & 21-03-2002   &$-$0.06  &   0.80   &   0.80   &  0.27  &   0.76   &   47 &   10 &\\
PKS 1145-071 A        & R  &  11 47 51.5   &  $-$07 24 41   &  1.342    & 02-05-2002   &$-$0.27  &   1.05   &   1.08   &  0.24  &   1.06   &   52 &    6 &\\
PKS 1145-071 B        & R  &  11 47 54.8   &  $-$07 24 44   &  1.345    & 21-03-2002   &$-$0.51  &$-$0.86   &   1.00   &  0.41  &   0.92   &  120 &   13 &\\
PKS 1145-071 B        & R  &  11 47 54.8   &  $-$07 24 44   &  1.345    & 02-05-2002   &$-$1.60  &$-$0.57   &   1.70   &  0.55  &   1.61   &  100 &   10 &\\
2QZ J114954+0012      & B  &  11 49 54.9   &  $+$00 12 55   &  1.596    & 20-03-2002     & 1.04  &$-$1.18   &   1.57   &  0.22  &   1.56   &  156 &    4 &\\
PKS 1148-171          & R  &  11 51 03.2   &  $-$17 24 00   &  1.751    & 23-03-2002     & 0.00  &   0.38   &   0.38   &  0.27  &   0.31   &   45 &   25 &\\
FIRST J1202+2631      & R  &  12 02 40.7   &  $+$26 31 38   &  0.478    & 22-03-2002     & 0.65  &$-$0.07   &   0.65   &  0.15  &   0.64   &  177 &    7 &\\
He 1202-0501          & U  &  12 04 53.0   &  $-$05 18 13   &  0.169    & 01-05-2002     & 0.16  &   0.36   &   0.39   &  0.14  &   0.37   &   33 &   11 &\\
PKS 1203-26           & R  &  12 05 33.2   &  $-$26 34 04   &  0.786    & 20-03-2002   &$-$0.59  &   0.62   &   0.86   &  0.20  &   0.83   &   67 &    7 &\\
SDSS J1206+0023       & B  &  12 06 27.6   &  $+$00 23 35   &  2.331    & 25-02-2003   &$-$0.38  &$-$0.86   &   0.94   &  0.15  &   0.93   &  123 &    5 &\\
PKS 1205-008          & R  &  12 07 41.7   &  $-$01 06 37   &  1.002    & 25-02-2003     & 0.03  &   0.19   &   0.19   &  0.14  &   0.15   &   41 &   26 &\\
SDSS J1208+0020       & B  &  12 08 34.8   &  $+$00 20 48   &  2.707    & 25-02-2003     & 0.02  &   0.26   &   0.26   &  0.14  &   0.23   &   43 &   17 &\\
He 1207-2118          & U  &  12 09 38.0   &  $-$21 34 49   &  0.457    & 22-03-2002   &$-$0.65  &$-$0.22   &   0.69   &  0.16  &   0.67   &   99 &    7 &\\
SDSS J1209-0023       & B  &  12 09 57.2   &  $-$00 23 02   &  1.860    & 22-03-2002     & 1.30  &$-$0.63   &   1.44   &  0.35  &   1.40   &  167 &    7 &\\
Q 1210+1507           & U  &  12 13 08.0   &  $+$14 51 06   &  1.613    & 01-05-2002     & 0.04  &   0.40   &   0.40   &  0.19  &   0.36   &   42 &   15 &\\
SDSS J1214-0001       & B  &  12 14 41.4   &  $-$00 01 38   &  1.041    & 20-03-2002   &$-$2.15  &$-$1.07   &   2.40   &  0.32  &   2.38   &  103 &    4 &\\
FIRST J1214+2803      & RB &  12 14 42.3   &  $+$28 03 29   &  0.698    & 23-03-2002   &$-$0.02  &$-$0.37   &   0.37   &  0.18  &   0.33   &  133 &   16 &\\

\hline
\end{tabular}
\begin{flushleft}
\end{flushleft}
\end{table*} 
\addtocounter{table}{-1}

\begin{table*}[htb]
\centering
\caption{continued}
\begin{tabular}{llccccrrrrrrrr}
\hline 
Object  &  & RA J2000 & DEC J2000 & $z$ & Date & $q$ \ \ & $u$ \ \ & $p$ \ \ & ${\sigma}_p$  & $p_0$ \ & $\theta$ \  & ${\sigma}_{\theta}$ &   \\
& & (h m s)& $(\degr$ $\arcmin$ $\arcsec$) & & dd-mm-yyyy & (\%)  & (\%) & (\%) & (\%) & (\%) &  ($\degr$) & ($\degr$) & \\
\hline \\

SDSS J1216+0107       & B  &  12 16 33.9   &  $+$01 07 33   &  2.017    & 20-03-2002   &$-$0.93  &   0.23   &   0.96   &  0.24  &   0.93   &   83 &    7 &\\
PKS 1215-002          & R  &  12 17 58.7   &  $-$00 29 47   &  0.420    & 21-04-2002   &$-$23.93  &$-$0.78   &  23.94   &  0.70  &  23.93  &   91 &   1 &\\
PKS 1216-010          & R  &  12 18 35.0   &  $-$01 19 54   &  0.415    & 02-05-2002   &$-$10.55  &$-$3.77   &  11.20   &  0.17  &  11.20  &  100 &   1 &\\
PG 1216+069           & R  &  12 19 20.9   &  $+$06 38 38   &  0.334    & 21-03-2002   &$-$0.60  &   0.07   &   0.60   &  0.13  &   0.59   &   87 &    6 &\\
Q 1217+1509           & U  &  12 19 46.4   &  $+$14 52 59   &  0.402    & 22-03-2002     & 0.45  &   0.26   &   0.52   &  0.26  &   0.46   &   15 &   16 &\\
PKS 1219+04           & R  &  12 22 22.5   &  $+$04 13 16   &  0.965    & 22-03-2002   &$-$3.10  &$-$4.62   &   5.56   &  0.15  &   5.56   &  118 &    1 &\\
Q 1221+1745           & R  &  12 23 34.9   &  $+$17 28 42   &  1.354    & 02-05-2002     & 0.50  &   0.64   &   0.81   &  0.19  &   0.79   &   26 &    7 &\\
4C 18.34              & R  &  12 23 46.2   &  $+$18 21 07   &  1.401    & 02-05-2002   &$-$0.07  &   0.11   &   0.13   &  0.27  &   0.00   &    -  &  -  &\\
PKS 1222+037          & R  &  12 24 52.3   &  $+$03 30 50   &  0.960    & 23-03-2002   &$-$0.89  &   0.43   &   0.99   &  0.27  &   0.95   &   77 &    8 &\\
PKS 1222+037          & R  &  12 24 52.3   &  $+$03 30 50   &  0.960    & 02-05-2002   &$-$2.40  &$-$0.73   &   2.51   &  0.22  &   2.50   &   98 &    2 &\\
PKS 1222+21           & R  &  12 24 54.5   &  $+$21 22 46   &  0.435    & 23-03-2002     & 1.36  &$-$0.68   &   1.52   &  0.13  &   1.51   &  167 &    3 &\\
RXS J12254-0418       & R  &  12 25 27.2   &  $-$04 18 57   &  0.137    & 01-05-2002     & 0.07  &   0.14   &   0.16   &  0.14  &   0.08   &   32 &   48 &\\
TON 1530              & R  &  12 25 27.4   &  $+$22 35 13   &  2.058    & 21-03-2002     & 0.86  &$-$0.34   &   0.92   &  0.14  &   0.91   &  169 &    4 &\\
SDSS J1227-0010       & B  &  12 27 26.9   &  $-$00 10 04   &  1.543    & 22-03-2002     & 0.48  &$-$0.40   &   0.62   &  0.28  &   0.57   &  160 &   14 &\\
SDSS J1230-0053       & B  &  12 30 56.6   &  $-$00 53 06   &  2.161    & 20-03-2002   &$-$0.10  &$-$0.53   &   0.54   &  0.24  &   0.49   &  130 &   14 &\\
SDSS J1231+0047       & B  &  12 31 24.7   &  $+$00 47 19   &  1.720    & 20-03-2002     & 0.22  &   1.44   &   1.46   &  0.40  &   1.40   &   41 &    8 &\\
SDSS J1234+0057       & B  &  12 34 27.8   &  $+$00 57 59   &  1.532    & 20-03-2002     & 1.35  &   0.09   &   1.35   &  0.23  &   1.33   &    2 &    5 &\\
FIRST J1240+2425      & R  &  12 40 09.1   &  $+$24 25 31   &  0.829    & 22-03-2002     & 0.30  &$-$0.12   &   0.32   &  0.16  &   0.29   &  169 &   16 &\\
Q 1244+1329           & U  &  12 46 54.0   &  $+$13 13 11   &  0.512    & 22-03-2002     & 0.01  &$-$0.14   &   0.14   &  0.14  &   0.03   &  137 &  120 &\\
Q 1244-0126           & U  &  12 47 30.9   &  $-$01 42 28   &  0.346    & 02-05-2002   &$-$0.55  &   0.98   &   1.12   &  0.36  &   1.07   &   60 &   10 &\\
PG 1254+047           & B  &  12 56 59.9   &  $+$04 27 34   &  1.018    & 21-04-2002     & 0.83  &$-$0.13   &   0.84   &  0.15  &   0.83   &  176 &    5 &\\
PKS 1256-229          & R  &  12 59 08.5   &  $-$23 10 38   &  1.365    & 22-03-2002     &15.40  &$-$16.16   &  22.32   &  0.15  &  22.32   &  157 &    1 &\\
SDSS J1300+0105       & B  &  13 00 58.1   &  $+$01 05 52   &  1.902    & 22-03-2002   &$-$0.45  &   0.56   &   0.72   &  0.31  &   0.66   &   64 &   13 &\\
SDSS J1301+0001       & B  &  13 01 36.1   &  $+$00 01 58   &  1.783    & 20-03-2002   &$-$0.22  &   0.20   &   0.30   &  0.17  &   0.26   &   69 &   19 &\\
SDSS J1302-0037       & B  &  13 02 08.2   &  $-$00 37 32   &  1.672    & 21-03-2002     & 0.45  &   1.29   &   1.37   &  0.20  &   1.35   &   35 &    4 &\\
SDSS J1305+0019       & B  &  13 05 06.7   &  $+$00 19 09   &  1.912    & 21-03-2002     & 0.30  &   0.78   &   0.84   &  0.28  &   0.79   &   34 &   10 &\\
EQS B1303+0205        & R  &  13 05 54.1   &  $+$01 49 30   &  0.740    & 23-03-2002     & 0.10  &   0.34   &   0.35   &  0.16  &   0.32   &   37 &   14 &\\
PKS 1303-250          & R  &  13 06 15.4   &  $-$25 17 20   &  0.738    & 22-03-2002   &$-$0.79  &$-$0.46   &   0.91   &  0.17  &   0.90   &  105 &    5 &\\
He 1304-1157          & R  &  13 07 03.9   &  $-$12 13 13   &  0.294    & 01-05-2002   &$-$0.69  &   0.47   &   0.83   &  0.18  &   0.82   &   73 &    6 &\\
He 1307-1651          & U  &  13 10 38.4   &  $-$17 07 02   &  1.173    & 01-05-2002   &$-$0.21  &   0.84   &   0.87   &  0.20  &   0.84   &   52 &    7 &\\
FIRST J1312+2319      & RB &  13 12 13.5   &  $+$23 19 58   &  1.508    & 21-03-2002     & 0.97  &$-$0.51   &   1.10   &  0.16  &   1.08   &  166 &    4 &\\
SDSS J1318+0022       & B  &  13 18 53.4   &  $+$00 22 11   &  2.078    & 21-03-2002   &$-$0.08  &   0.05   &   0.09   &  0.24  &   0.00   &  -    &  -   &\\
PKS 1317-00           & R  &  13 19 38.7   &  $-$00 49 40   &  0.892    & 23-03-2002     & 0.23  &   0.06   &   0.24   &  0.20  &   0.16   &    7 &   37 &\\
SDSS J1323-0038       & B  &  13 23 04.6   &  $-$00 38 57   &  1.827    & 22-03-2002     & 0.97  &   0.57   &   1.13   &  0.21  &   1.11   &   15 &    5 &\\
SDSS J1327+0035       & B  &  13 27 42.9   &  $+$00 35 33   &  1.876    & 23-03-2002   &$-$0.75  &   0.82   &   1.11   &  0.23  &   1.09   &   66 &    6 &\\
PKS 1327-21           & R  &  13 30 07.1   &  $-$21 42 02   &  0.525    & 23-03-2002     & 0.50  &   0.15   &   0.52   &  0.15  &   0.50   &    8 &    8 &\\
PKS 1328-264          & R  &  13 31 11.7   &  $-$26 39 09   &  0.883    & 22-03-2002   &$-$0.09  &   0.00   &   0.09   &  0.18  &   0.00   &  -    &  -   &\\
PKS 1328-173          & R  &  13 31 35.9   &  $-$17 36 34   &  0.329    & 02-05-2002   &$-$0.16  &   0.31   &   0.35   &  0.18  &   0.31   &   59 &   17 &\\
PKS 1335+023          & R  &  13 37 39.6   &  $+$02 06 57   &  1.356    & 23-03-2002     & 0.54  &   0.43   &   0.69   &  0.20  &   0.66   &   19 &    8 &\\
PKS 1335-06           & R  &  13 38 08.0   &  $-$06 27 11   &  0.625    & 02-05-2002   &$-$0.74  &$-$0.03   &   0.74   &  0.25  &   0.70   &   91 &   10 &\\
CTS J13.07            & U  &  13 42 04.4   &  $-$18 18 01   &  2.210    & 21-03-2002     & 0.64  &   0.53   &   0.83   &  0.15  &   0.82   &   20 &    5 &\\
UM 607                & U  &  13 42 51.7   &  $-$00 53 45   &  0.326    & 02-05-2002   &$-$0.15  &   0.15   &   0.21   &  0.14  &   0.17   &   67 &   23 &\\
Q 1359-058            & U  &  14 01 41.1   &  $-$06 08 23   &  1.996    & 21-03-2002   &$-$0.63  &$-$0.25   &   0.68   &  0.16  &   0.66   &  101 &    7 &\\
FIRST J1408+3054      & RB &  14 08 06.2   &  $+$30 54 48   &  0.842    & 20-03-2002     & 0.15  &$-$0.25   &   0.29   &  0.17  &   0.25   &  150 &   19 &\\
SDSS J1409+0048       & B  &  14 09 18.7   &  $+$00 48 24   &  1.999    & 21-03-2002     & 1.95  &   3.39   &   3.91   &  0.28  &   3.90   &   30 &    2 &\\
TEX 1414+235          & R  &  14 17 17.9   &  $+$23 17 20   &  0.921    & 02-05-2002   &$-$0.14  &   0.08   &   0.16   &  0.18  &   0.00   &   -  &   -  &\\
HS 1417+2547          & RB &  14 20 13.1   &  $+$25 34 04   &  2.200    & 20-03-2002   &$-$0.68  &$-$0.78   &   1.03   &  0.18  &   1.02   &  114 &    5 &\small{$\star\star$}\\
FIRST J1427+2709      & B  &  14 27 03.6   &  $+$27 09 40   &  1.170    & 20-03-2002   &$-$1.27  &   0.46   &   1.35   &  0.25  &   1.33   &   80 &    5 &\\
PG 1435-067           & U  &  14 38 16.2   &  $-$06 58 20   &  0.129    & 01-05-2002     & 0.24  &   1.03   &   1.06   &  0.14  &   1.05   &   38 &    4 &\small{$\star\star$}\\
2MA 1519+1838         & U  &  15 19 01.5   &  $+$18 38 04   &  0.187    & 20-03-2002   &$-$0.17  &$-$1.47   &   1.48   &  0.22  &   1.46   &  132 &    4 &\\
2MA 1543+1937         & U  &  15 43 07.7   &  $+$19 37 51   &  0.228    & 20-03-2002     & 0.20  &   1.93   &   1.94   &  0.17  &   1.93   &   42 &    3 &\\
3C 323.1              & R  &  15 47 43.6   &  $+$20 52 16   &  0.266    & 01-05-2002     & 0.93  &   0.68   &   1.15   &  0.13  &   1.14   &   18 &    3 &\\
MARK 877              & R  &  16 20 11.3   &  $+$17 24 28   &  0.114    & 02-05-2002   &$-$0.66  &   0.14   &   0.67   &  0.13  &   0.66   &   84 &    6 &\\
MARK 877              & R  &  16 20 11.3   &  $+$17 24 28   &  0.114    & 23-03-2002   &$-$0.56  &   0.13   &   0.57   &  0.13  &   0.56   &   83 &    7 &\\
RXS J16212+1819       & R  &  16 21 14.4   &  $+$18 19 49   &  0.125    & 01-05-2002   &$-$0.44  &$-$0.30   &   0.53   &  0.16  &   0.51   &  107 &    9 &\\
2MA J1714+2602	      & U  &  17 14 42.7   &  $+$26 02 48   &  0.163    & 01-05-2002   &$-$0.27  &$-$0.13   &   0.30   &  0.18  &   0.26   &  103 &   20 &\\
FIRST J21079-0620      & RB&  21 07 57.7 & $-$06 20 10 & 0.644  &  19-10-2003 &     0.44 &  $-$1.03 &      1.12 &     0.22 &     1.10 &     147 &     6 & \\
Q 2116-4439            & B &  21 20 11.6 & $-$44 26 54 & 1.480  &  27-08-2000 &     0.53 &     0.22 &      0.57 &     0.25 &     0.52 &      11 &    13 & \\ 

\hline
\end{tabular}
\end{table*} 
\addtocounter{table}{-1}

\begin{table*}[htb]
\centering
\caption{continued}
\begin{tabular}{llccccrrrrrrrr}
\hline 
Object  &  & RA J2000 & DEC J2000 & $z$ & Date & $q$ \ \ & $u$ \ \ & $p$ \ \ & ${\sigma}_p$  & $p_0$ \ & $\theta$ \  & ${\sigma}_{\theta}$ &   \\
& & (h m s)& $(\degr$ $\arcmin$ $\arcsec$) & & dd-mm-yyyy & (\%)  & (\%) & (\%) & (\%) & (\%) &  ($\degr$) & ($\degr$) & \\
\hline \\

PKS 2124-12            & R &  21 27 25.9 & $-$11 51 17 & 0.873  &  21-08-2003 &     0.04 &     0.24 &      0.24 &     0.43 &     0.00 &       - &     - & \\
SDSS J2131-0839        & B &  21 31 13.9 & $-$08 39 13 & 1.983  &  20-10-2003 &     0.58 &  $-$0.20 &      0.61 &     0.27 &     0.55 &     171 &    14 & \\
PKS 2128-315           & R &  21 31 23.2 & $-$31 21 13 & 0.990  &  27-08-2000 &  $-$0.08 &     0.00 &      0.08 &     0.22 &     0.00 &      -  &    -  & \\ 
Q 2128-4327            & U &  21 31 25.3 & $-$43 13 52 & 0.920  &  20-08-2003 &     0.08 &  $-$0.40 &      0.41 &     0.28 &     0.34 &     141 &    24 & \\
SDSS J2131-0700        & B &  21 31 38.9 & $-$07 00 13 & 2.048  &  20-08-2003 &     0.08 &     1.78 &      1.78 &     0.32 &     1.75 &      44 &     5 & \\
PHL 1631              & R  &  21 35 13.1   &  $-$00 52 44   &  1.660    & 01-05-2002     &$-$0.57  &  $-$0.60   &   0.83   &  0.25  &   0.79   &  113 &    9 &\\
FIRST J21378+0012     & R  &  21 37 48.4   &  $+$00 12 20   &  1.666    & 02-05-2002     &$-$0.22  &  $-$0.42   &   0.47   &  0.25  &   0.42   &  121 &   18 &\\
FIRST J21378+0012     & R  &  21 37 48.4   &  $+$00 12 20   &  1.666    & 01-05-2002     &   0.24  &  $-$0.36   &   0.43   &  0.23  &   0.38   &  152 &   17 &\\
PKS 2135-248           & R &  21 38 37.2 & $-$24 39 55 & 0.819  &  27-08-2000 &     0.58 &  $-$0.21 &      0.62 &     0.35 &     0.53 &     170 &    19 & \\ 
PKS 2138-377           & R &  21 41 52.4 & $-$37 29 12 & 0.425  &  20-08-2003 &     0.43 &     0.35 &      0.56 &     0.28 &     0.50 &      19 &    16 & \\
FIRST J21424-0821      & R &  21 42 25.4 & $-$08 21 22 & 0.570  &  20-10-2003 &     0.61 &  $-$0.51 &      0.79 &     0.22 &     0.76 &     160 &     8 & \\
2E 2141+0402           & U &  21 44 23.0 & $+$04 16 27 & 0.463  &  22-08-2003 &  $-$0.63 &  $-$0.55 &      0.84 &     0.25 &     0.80 &     111 &     9 & \\
CT 677                 & U &  21 44 40.8 & $-$49 21 22 & 1.440  &  21-08-2003 &  $-$0.10 &  $-$0.62 &      0.63 &     0.25 &     0.58 &     131 &    12 & \\
PKS 2144-362           & R &  21 47 31.1 & $-$36 01 52 & 2.081  &  21-08-2003 &  $-$0.04 &     0.66 &      0.66 &     0.28 &     0.60 &      46 &    13 & \\
HS 2146+0428           & U &  21 48 32.2 & $+$04 42 16 & 1.320  &  22-08-2003 &  $-$0.42 &  $-$0.42 &      0.59 &     0.27 &     0.53 &     113 &    15 & \\
PKS 2149-20            & R &  21 51 51.1 & $-$19 46 06 & 0.424  &  20-08-2003 &     1.10 &     2.01 &      2.29 &     0.31 &     2.27 &      31 &     4 & \\
PKS 2149-306           & R &  21 51 55.4 & $-$30 27 54 & 2.345  &  22-08-2003 &  $-$0.18 &     0.23 &      0.29 &     0.28 &     0.10 &      64 &    77 & \\
PKS 2150+05            & R &  21 53 24.5 & $+$05 36 19 & 1.980  &  20-10-2003 &  $-$0.22 &     0.22 &      0.31 &     0.25 &     0.22 &      67 &    32 & \\
PKS 2151-153           & R &  21 54 07.5 & $-$15 01 30 & 1.208  &  20-08-2003 &     0.12 &     0.21 &      0.24 &     0.26 &     0.00 &       - &     - & \\
PKS 2153-204           & R &  21 56 33.8 & $-$20 12 30 & 1.309  &  27-08-2000 &  $-$0.20 &     0.01 &      0.20 &     0.21 &     0.00 &      -  &    -  & \\ 
PB 5020                & U &  21 58 19.1 & $+$02 08 49 & 0.560  &  20-10-2003 &     0.04 &  $-$0.51 &      0.51 &     0.22 &     0.47 &     137 &    14 & \\
RXS J21594+0113        & R &  21 59 24.1 & $+$01 13 05 & 1.000  &  22-08-2003 &  $-$0.17 &     0.21 &      0.26 &     0.26 &     0.00 &       - &     - & \\
PKS 2157-200           & R &  22 00 07.8 & $-$19 45 46 & 1.198  &  28-08-2000 &  $-$0.12 &  $-$0.06 &      0.13 &     0.24 &     0.00 &      -  &    -  & \\ 
CTS A09.14             & R &  22 01 51.2 & $-$33 44 37 & 2.210  &  02-05-2002 &  $-$0.05 &     0.47 &      0.47 &     0.16 &     0.45 &      48 &    10 &\\
Q 2200-1816            & U &  22 03 11.6 & $-$18 01 42 & 1.160  &  28-08-2000 &     0.07 &  $-$0.03 &      0.08 &     0.21 &     0.00 &      -  &    -  & \\ 
CTS H10.02             & U &  22 05 31.3 & $-$54 05 41 & 1.150  &  22-08-2003 &     0.33 &     0.01 &      0.33 &     0.28 &     0.22 &       0 &    36 & \\
PKS 2203-18            & R &  22 06 10.5 & $-$18 35 39 & 0.619  &  22-08-2003 &     0.58 &     1.12 &      1.26 &     0.29 &     1.23 &      31 &     7 & \\
PKS 2203-215           & R &  22 06 41.4 & $-$21 19 42 & 0.577  &  22-08-2003 &  $-$0.06 &     0.99 &      0.99 &     0.30 &     0.95 &      47 &     9 & \\
PKS 2204-54            & R &  22 07 43.6 & $-$53 46 33 & 1.206  &  20-08-2003 &  $-$0.33 &  $-$1.78 &      1.81 &     0.26 &     1.79 &     130 &     4 & \\
Q 2207-1627            & U &  22 10 34.7 & $-$16 12 19 & 1.296  &  28-08-2000 &  $-$0.32 &  $-$0.05 &      0.32 &     0.27 &     0.23 &      94 &    34 & \\ 
Q 2208-1720            & B &  22 11 15.4 & $-$17 05 26 & 1.210  &  27-08-2000 &     0.45 &  $-$0.89 &      1.00 &     0.24 &     0.97 &     148 &     7 & \\ 
Q 2208-1720            & B &  22 11 15.4 & $-$17 05 26 & 1.210  &  19-10-2003 &     0.38 &  $-$0.79 &      0.87 &     0.24 &     0.84 &     148 &     8 & \\
CXOU J221257-2221      & U &  22 12 57.4 & $-$22 21 35 & 1.167  &  22-08-2003 &  $-$0.11 &     0.04 &      0.11 &     0.34 &     0.00 &       - &     - & \\
PKS 2213-283           & R &  22 16 00.0 & $-$28 03 30 & 0.946  &  28-08-2000 &  $-$0.81 &  $-$0.23 &      0.84 &     0.23 &     0.81 &      98 &     8 & \\ 
MS 22152-0347          & R &  22 17 47.9 & $-$03 32 39 & 0.241  &  28-08-2000 &  $-$0.49 &     0.08 &      0.50 &     0.27 &     0.43 &      85 &    18 & \\ 
PKS 2215-508           & R &  22 18 19.1 & $-$50 38 42 & 1.356  &  27-08-2000 &     0.69 &  $-$0.42 &      0.81 &     0.22 &     0.78 &     164 &     8 & \\ 
FIRST 22186-0853       & R &  22 18 41.9 & $-$08 53 05 & 0.750  &  20-08-2003 &     0.71 &     0.03 &      0.72 &     0.31 &     0.66 &       1 &    14 &$\star$\\
HS 2217+0451           & U &  22 19 57.6 & $+$05 06 08 & 1.380  &  22-08-2003 &  $-$0.24 &  $-$0.75 &      0.79 &     0.27 &     0.75 &     126 &    10 &$\star\star$\\
RXS J22208-3156        & U &  22 20 49.7 & $-$31 56 54 & 0.503  &  19-10-2003 &  $-$0.25 &  $-$0.49 &      0.55 &     0.28 &     0.49 &     121 &    16 & \\
RX J22217-3327         & U &  22 21 42.9 & $-$33 27 28 & 0.562  &  19-10-2003 &  $-$0.19 &     0.13 &      0.23 &     0.23 &     0.00 &       - &     - & \\
Q 2220-3851            & U &  22 23 53.2 & $-$38 36 11 & 0.830  &  19-10-2003 &     0.01 &  $-$0.05 &      0.05 &     0.23 &     0.00 &       - &     - & \\
CTS A10.09             & U &  22 29 02.3 & $-$33 15 03 & 0.868  &  28-08-2000 &  $-$0.36 &  $-$0.39 &      0.53 &     0.21 &     0.49 &     114 &    12 & \\ 
PKS 2226-41            & R &  22 29 18.6 & $-$40 51 33 & 0.446  &  20-08-2003 &  $-$0.35 &     0.75 &      0.82 &     0.32 &     0.76 &      57 &    12 & \\
RXS J22307+0309        & U &  22 30 45.9 & $+$03 09 19 & 1.321  &  22-08-2003 &  $-$0.34 &  $-$0.69 &      0.77 &     0.27 &     0.72 &     122 &    11 &$\star\star$\\
PKS 2227-445           & R &  22 30 56.5 & $-$44 16 30 & 1.326  &  22-08-2003 &     4.28 &     3.05 &      5.26 &     0.48 &     5.24 &      18 &     3 & \\
PKS 2232-488           & R &  22 35 13.3 & $-$48 35 59 & 0.510  &  22-08-2003 &     3.42 &     1.29 &      3.66 &     0.26 &     3.65 &      10 &     2 & \\
UM 657                 & U &  22 41 25.2 & $-$17 14 25 & 1.360  &  22-08-2003 &     0.28 &     0.22 &      0.36 &     0.31 &     0.23 &      19 &    38 & \\
RXS J22418-4405        & U &  22 41 55.3 & $-$44 04 58 & 0.545  &  19-10-2003 &     0.09 &     0.09 &      0.13 &     0.22 &     0.00 &       - &     - & \\
PKS 2240-260           & R &  22 43 26.4 & $-$25 44 30 & 0.774  &  20-10-2003 &  $-$2.11 & $-$14.63 &     14.78 &     0.21 &    14.78 &     131 &     1 & \\
PKS 2243-03            & R &  22 46 11.3 & $-$03 00 39 & 1.348  &  28-08-2000 &     0.23 &     0.04 &      0.23 &     0.23 &     0.00 &      -  &    -  & \\ 
PKS 2246-309           & R &  22 49 19.0 & $-$30 39 13 & 1.307  &  27-08-2000 &     0.11 &     0.17 &      0.20 &     0.29 &     0.00 &      -  &    -  & \\ 
Q 2247+0135            & U &  22 49 35.6 & $+$01 51 48 & 1.128  &  20-10-2003 &  $-$1.07 &     0.29 &      1.11 &     0.25 &     1.08 &      82 &     7 & \\
PKS 2247+13            & R &  22 49 45.0 & $+$13 31 10 & 0.767  &  28-08-2000 &     0.14 &     0.30 &      0.33 &     0.28 &     0.22 &      32 &    36 & \\ 
FIRST J22541+0056      & R &  22 54 09.6 & $+$00 56 28 & 1.150  &  20-08-2003 &  $-$0.20 &  $-$0.86 &      0.89 &     0.26 &     0.85 &     129 &     9 & \\
PB 7348                & U &  22 55 38.6 & $-$11 14 53 & 1.330  &  20-10-2003 &  $-$0.14 &  $-$0.80 &      0.81 &     0.23 &     0.78 &     130 &     8 & \\
Q 2254-2447            & U &  22 57 05.9 & $-$24 31 22 & 1.091  &  27-08-2000 &  $-$0.01 &  $-$0.13 &      0.13 &     0.21 &     0.00 &      -  &    -  & \\ 
PKS 2257-270           & R &  23 00 25.5 & $-$26 44 23 & 1.476  &  28-08-2000 &  $-$0.17 &     0.11 &      0.20 &     0.37 &     0.00 &      -  &    -  & \\ 

\hline
\end{tabular}
\end{table*} 
\addtocounter{table}{-1}

\begin{table*}[htb]
\centering
\caption{continued}
\begin{tabular}{llccccrrrrrrrr}
\hline 
Object  &  & RA J2000 & DEC J2000 & $z$ & Date & $q$ \ \ & $u$ \ \ & $p$ \ \ & ${\sigma}_p$  & $p_0$ \ & $\theta$ \  & ${\sigma}_{\theta}$ &   \\
& & (h m s)& $(\degr$ $\arcmin$ $\arcsec$) & & dd-mm-yyyy & (\%)  & (\%) & (\%) & (\%) & (\%) &  ($\degr$) & ($\degr$) & \\
\hline \\

PKS 2301+06            & R &  23 04 28.3 & $+$06 20 08 & 1.268  &  28-08-2000 &     3.03 &  $-$2.11 &      3.69 &     0.26 &     3.68 &     163 &     2 & \\ 
PKS 2302-279           & R &  23 05 15.3 & $-$27 38 41 & 1.435  &  28-08-2000 &     0.78 &     0.26 &      0.82 &     0.21 &     0.80 &       9 &     7 & \\ 
PKS 2303-052           & R &  23 06 15.3 & $-$04 59 49 & 1.139  &  19-10-2003 &  $-$0.21 &  $-$0.28 &      0.35 &     0.24 &     0.29 &     116 &    24 & \\
PKS 2312-319           & R &  23 14 48.5 & $-$31 38 39 & 1.323  &  27-08-2000 &  $-$0.40 &  $-$0.02 &      0.40 &     0.23 &     0.35 &      91 &    19 & \\ 
SDSS J2319-0024        & B &  23 19 58.7 & $-$00 24 49 & 1.889  &  20-08-2003 &     1.58 &  $-$0.95 &      1.85 &     0.30 &     1.83 &     164 &     5 & \\
RXS J23202-4700        & U &  23 20 16.6 & $-$47 00 51 & 0.430  &  19-10-2003 &  $-$0.15 &     0.16 &      0.22 &     0.25 &     0.00 &       - &     - & \\
CTS A13.08             & U &  23 22 10.8 & $-$34 47 57 & 0.420  &  20-10-2003 &  $-$0.01 &  $-$0.32 &      0.32 &     0.22 &     0.26 &     134 &    24 & \\
PKS 2320-035           & R &  23 23 32.0 & $-$03 17 05 & 1.411  &  28-08-2000 &  $-$9.56 &     0.15 &      9.56 &     0.20 &     9.56 &      90 &     1 & \\ 
PKS 2326-502           & R &  23 29 20.9 & $-$49 55 41 & 0.518  &  19-10-2003 &     3.29 &  $-$2.14 &      3.92 &     0.33 &     3.91 &     164 &     2 & \\
RXS J23300-5506        & U &  23 30 01.8 & $-$55 06 23 & 0.494  &  20-10-2003 &     0.12 &     0.07 &      0.14 &     0.23 &     0.00 &       - &     - & \\
PKS 2332-017           & R &  23 35 20.4 & $-$01 31 09 & 1.184  &  28-08-2000 &  $-$4.85 &  $-$0.37 &      4.86 &     0.19 &     4.86 &      92 &     1 & \\ 
FIRST 23361-0954       & RB&  23 36 11.7 & $-$09 54 27 & 1.760  &  20-08-2003 &     0.75 &  $-$0.64 &      0.99 &     0.34 &     0.93 &     160 &    10 & \\
HS 2334+1716           & U &  23 36 41.1 & $+$17 32 50 & 0.470  &  20-08-2003 &  $-$0.72 &  $-$0.17 &      0.74 &     0.40 &     0.65 &      97 &    18 & \\
HE 2335-3029           & U &  23 37 42.3 & $-$30 13 17 & 1.119  &  20-08-2003 &     0.00 &  $-$0.12 &      0.12 &     0.26 &     0.00 &       - &     - & \\
PKS 2335-18            & R &  23 37 56.7 & $-$17 52 21 & 1.450  &  27-08-2000 &     0.10 &  $-$0.14 &      0.17 &     0.22 &     0.00 &      -  &    -  & \\ 
PKS 2335-027           & R &  23 37 57.3 & $-$02 30 58 & 1.072  &  20-08-2003 &  $-$2.72 &  $-$2.28 &      3.55 &     0.30 &     3.54 &     110 &     2 & \\
FIRST 23382-0140       & R &  23 38 14.7 & $-$01 40 08 & 1.060  &  20-08-2003 &     0.29 &  $-$0.44 &      0.53 &     0.30 &     0.46 &     152 &    19 & \\
PKS 2338-290           & R &  23 40 51.1 & $-$28 48 42 & 0.446  &  19-10-2003 &     0.41 &     0.24 &      0.48 &     0.27 &     0.42 &      15 &    19 & \\
RXS J23432-3638        & R &  23 43 13.5 & $-$36 37 54 & 0.622  &  19-10-2003 &     0.17 &  $-$0.10 &      0.19 &     0.22 &     0.00 &       - &     - & \\
UM 173                 & U &  23 43 26.6 & $-$00 03 16 & 1.366  &  27-08-2000 &  $-$0.35 &  $-$0.29 &      0.45 &     0.26 &     0.39 &     110 &    19 & \\ 
FIRST J23440+0038      & R &  23 44 03.1 & $+$00 38 03 & 1.230  &  20-10-2003 &  $-$0.16 &  $-$0.29 &      0.33 &     0.23 &     0.27 &     120 &    25 & \\
Q 2344+18     	       & U &  23 47 25.7 & $+$18 44 50 & 0.138  &  27-08-2000 &  $-$1.01 &     0.06 &      1.01 &     0.32 &     0.96 &      88 &    10 & \\ 
SDSS J2348+0029        & B &  23 48 12.4 & $+$00 29 40 & 1.946  &  20-10-2003 &  $-$0.02 &  $-$0.91 &      0.91 &     0.30 &     0.86 &     134 &    10 & \\
HE 2346-3635           & U &  23 48 44.8 & $-$36 18 27 & 0.541  &  19-10-2003 &     0.33 &     0.54 &      0.64 &     0.25 &     0.59 &      29 &    12 & \\
FIRST 23501-1017       & R &  23 50 08.3 & $-$10 17 35 & 1.310  &  20-10-2003 &  $-$0.89 &  $-$0.57 &      1.05 &     0.29 &     1.01 &     106 &     8 &$\star$\\
PKS 2348-252           & R &  23 50 49.8 & $-$24 57 04 & 1.386  &  27-08-2000 &     0.21 &  $-$0.23 &      0.31 &     0.22 &     0.25 &     156 &    25 & \\ 
SDSS J2352+0105        & B &  23 52 38.1 & $+$01 05 52 & 2.156  &  22-08-2003 &     0.92 &     1.30 &      1.59 &     0.26 &     1.57 &      27 &     5 &R\\
FIRST 23532-1041       & R &  23 53 14.3 & $-$10 41 05 & 1.240  &  20-08-2003 &     0.09 &  $-$0.26 &      0.28 &     0.27 &     0.10 &     145 &    76 & \\
UM 186                 & U &  23 53 21.8 & $-$01 15 29 & 0.993  &  27-08-2000 &     0.06 &  $-$0.48 &      0.48 &     0.21 &     0.44 &     139 &    14 & \\ 
PKS 2351-006           & R &  23 54 09.2 & $-$00 19 48 & 0.463  &  20-08-2003 &     0.35 &     0.15 &      0.38 &     0.27 &     0.30 &      11 &    25 & \\
SDSS J2356-0036        & B &  23 56 28.9 & $-$00 36 02 & 2.936  &  20-08-2003 &     1.52 &     0.98 &      1.81 &     0.34 &     1.78 &      16 &     5 &$i$\\
PKS 2354+14            & R &  23 57 18.4 & $+$14 46 10 & 1.816  &  20-10-2003 &     0.11 &     0.45 &      0.47 &     0.36 &     0.36 &      38 &    29 & \\
PB 5630                & U &  23 57 29.3 & $+$00 29 40 & 0.410  &  20-10-2003 &  $-$0.57 &     0.34 &      0.67 &     0.30 &     0.61 &      74 &    14 &$\star$\\
PKS 2356+196           & R &  23 58 46.1 & $+$19 55 21 & 1.066  &  20-08-2003 &  $-$0.54 &     0.39 &      0.66 &     0.28 &     0.60 &      72 &    13 &$\star\star$\\
SDSS J2358-0024        & B &  23 58 59.5 & $-$00 24 26 & 1.757  &  20-08-2003 &     1.03 &  $-$1.02 &      1.46 &     0.33 &     1.42 &     158 &     7 & \\
QSO J2359-12           & B &  23 59 53.6 & $-$12 41 49 & 0.868  &  27-08-2000 &     2.15 &  $-$3.51 &      4.12 &     0.20 &     4.11 &     151 &     1 & \\ 
SDSS J2359+0051        & B &  23 59 58.2 & $+$00 51 40 & 2.023  &  22-08-2003 &  $-$0.42 &     0.08 &      0.42 &     0.39 &     0.21 &      85 &    54 & \\
\hline
\end{tabular}
\begin{flushleft}
{\footnotesize Notes : $\star\star$ in the last column marks
polarization measurements most probably contaminated by interstellar
polarization, while $\star$ marks objects possibly contaminated.  R
and $i$ indicate observations done in filters other than V.}
\end{flushleft}
\end{table*}